\documentclass[twocolumn,aps,prl]{revtex4-1}

\usepackage[usenames,dvipsnames]{xcolor}
\usepackage{amsmath}
\usepackage{amsfonts}
\usepackage{pst-node}
\usepackage{tikz-cd} 
\usepackage{multirow}

\usepackage{amssymb}
\usepackage[colorlinks=true,citecolor=blue,linkcolor=red]{hyperref}
\usepackage{graphicx}
\usepackage{bbold}					
\usepackage[makeroom]{cancel}		
\usepackage{multirow}				
\usepackage[normalem]{ulem}        
\usepackage{array}
\usepackage{pdfpages}
	\makeatletter
	\AtBeginDocument{\let\LS@rot\@undefined}
	\makeatother
	
	\newcolumntype{x}[1]{>{\centering\let\newline\\\arraybackslash\hspace{0pt}}p{#1}}

	\DeclareMathOperator{\diag}{diag}  	

	\DeclareMathAlphabet{\mathbbold}{U}{bbold}{m}{n}

	\newcounter{subeqn} %

	\makeatletter
	\@addtoreset{subeqn}{equation}
	\makeatother

\definecolor{TB}{rgb}{0,0,0} 

\begin{document}

\title{Non-perturbative Breakdown of Bloch's Theorem and Hermitian Skin Effects}

\author{Zhesen Yang}\email[Corresponding author: ]{yangzs@iphy.ac.cn}

\affiliation{Kavli Institute for Theoretical Sciences, University of Chinese Academy of Sciences, Beijing 100190, China}

\date{\today}
	
\begin{abstract}
In conventional Hermitian systems with the open boundary condition, Bloch's theorem is perturbatively broken down, which means although the crystal momentum is not a good quantum number, the eigenstates are the superposition of several extended Bloch waves. In this paper, we show that Bloch's theorem can be non-perturbatively broken down in some Hermitian Bosonic systems. The quasiparticles of the system are the superposition of localized non-Bloch waves, which are characterized by the complex momentum whose imaginary part determines the localization properties. Our work is a Hermitian generalization of the non-Hermitian skin effect, although they share the same mechanism.  
\end{abstract} 
	
\maketitle

{\em Introduction}---Bloch's theorem, which is one of the cornerstones in condensed matter physics, plays a fundamental role in the development of many theories, like band theory, Fermi liquid theory, and BCS theory~\cite{Bloch_theorem,textbook1,Ashcroft76,coleman_2015,Pines_1999,PhysRev.108.1175}. It states that when the system has (discrete) translational symmetry, the corresponding eigenstates can be labeled by a conserved quantity, or good quantum number---(crystal) momentum. Strictly speaking, Bloch's theorem only applies to the system with periodic boundary condition (PBC), or with infinity boundary condition (IBC)~\footnote{means the wavefunction must be bounded at the infinity $\pm\infty$}. Based on Bloch's theorem, many important quantities can be expressed as the integral over the entire Brillouin zone (BZ) or over the Fermi surface~\cite{mah00,altland_simons_2010,10.2307/j.ctt19cc2gc}. 

\begin{figure}[t]
	\centerline{\includegraphics[width=1\linewidth]{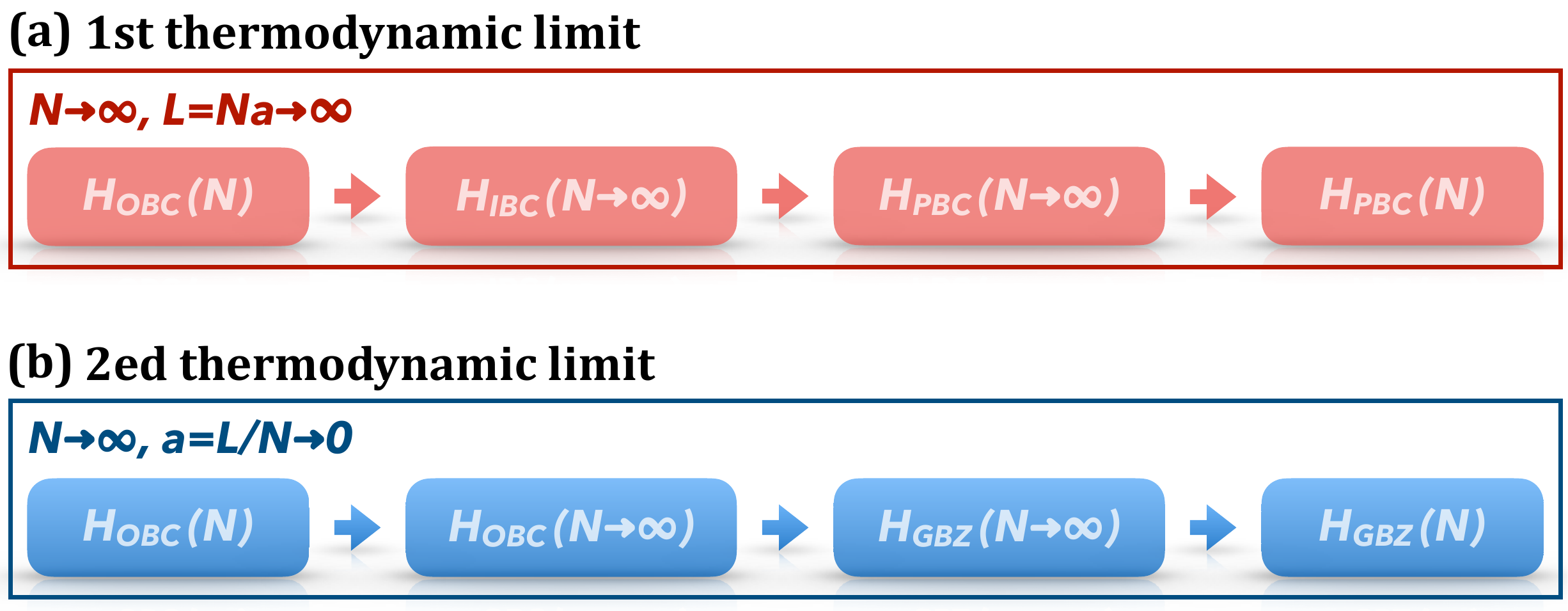}}
	\caption{Two types of thermodynamic limit. The second one in (b) gives the correct asymptotic solution of the finite size OBC Hamiltonian, since it preserves the boundary condition. 
		\label{F1}}
\end{figure}

In realistic macroscopic materials, the system is neither with PBC nor with IBC. Furthermore, the translational symmetry is explicitly broken down due to the existence of boundaries. It is natural to ask why can we still use Bloch's theorem to understand the physical properties of real materials? In traditional textbooks of solid-state physics, a thermodynamic limit argument is provided to explain the above question~\cite{Ashcroft76,born1988dynamical}. As shown in Fig.~\ref{F1} (a), since the lattice size of the macroscopic material is very large, e.g. $N\simeq 10^{23}$, its asymptotic behaviors can be described by the thermodynamic limit $N\rightarrow\infty$. Therefore, if we fix the lattice constant $a$, the system length will extend to infinity, i.e. $L=aN\rightarrow \infty$, which implies the boundary condition has been replaced by the IBC. Since the wave function should be bounded at the infinity $x\rightarrow\pm \infty$, the corresponding momentum is restricted to be real numbers. This revives Bloch's theorem. Therefore,  for a finite-size system with the open boundary condition (OBC), it can be approximately described by the same system with PBC. Physically, this can be understood from the fact that the boundary only acts as the role of scattering potential. The truth eigenstate of the OBC Hamiltonian is a superposition of these scattered Bloch waves with the same energy. In this sense, one can still use Bloch’s theorem to understand real materials with the OBC. Through the above argument, one can notice that, although the existence of boundary definitely breaks the translational symmetry, Bloch’s theorem is perturbatively broken down.

Even for most uniform interacting systems, Bloch's theorem is also approximately preserved. Indeed, in a many-body system, the elementary excitations (or quasiparticles) are referred to the eigenmodes with well-defined dispersion relation~\cite{anderson1997concepts,zhai2021ultracold}. Although the effect of many-boy interaction often corrects some physical quantities (for example the excitation energy and effective mass), the concept of momentum is believed to be preserved and well-defined. This means the excitations of the system are extended states in the bulk, which is consistent with the physical intuition in a disorder free system.

\begin{figure*}[t]
	\centerline{\includegraphics[width=1\linewidth]{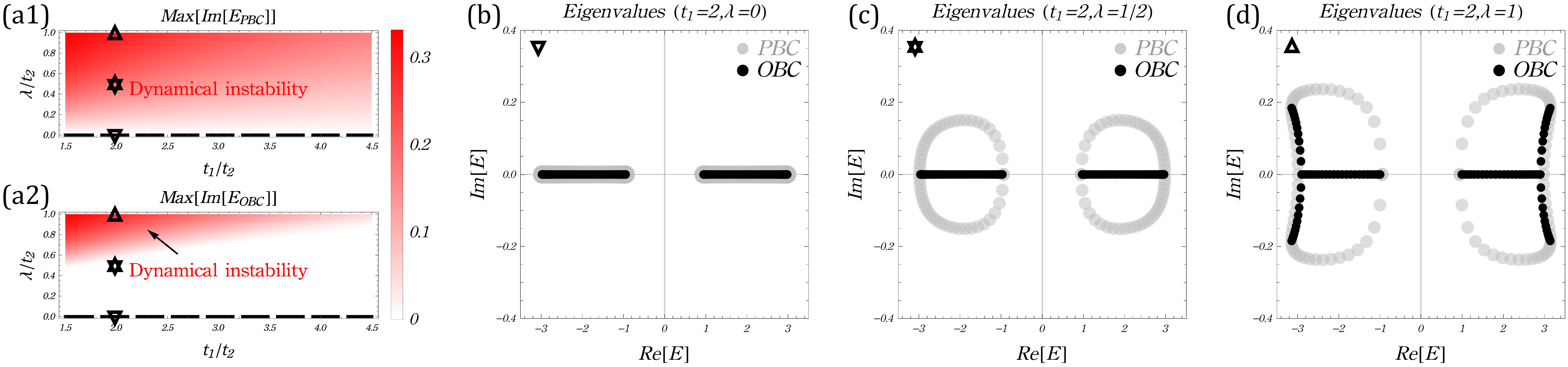}}
	\caption{The non-perturbative breakdown of Bloch's theorem. (a) shows the dynamical instability regions with PBC and OBC. (b)-(d) show the corresponding PBC and OBC spectrum. Here the parameters are chosen as $t_2=1$ and $\Delta=1/3$. 
		\label{F2}}
\end{figure*}

In this paper, we show that all the above arguments are challenged in some {\em Hermitian} Bosonic systems, in which all the eigenstates (or quasiparticles) are localized at the boundary, indicating that the Bloch’s theorem is nonperturbatively broken down. This phenomenon generalizes the concept of skin effect from non-Hermitian~\cite{yaoEdgeStatesTopological2018b,yaoNonHermitianChernBands2018b,PhysRevLett.123.170401,PhysRevLett.123.246801,xue2020nonhermitian,Xiao:2020aa,yokomizoNonBlochBandTheory2019a,PhysRevResearch.2.043045,PhysRevLett.121.026808,PhysRevResearch.2.043046,bergholtz2020exceptional,yang2020exceptional,PhysRevLett.125.126402,PhysRevLett.125.226402,PhysRevLett.125.186802,PhysRevB.102.045412,PhysRevResearch.2.043167,PhysRevX.8.031079,PhysRevX.9.041015,PhysRevLett.123.097701,okumaTopologicalOriginNonHermitian2019,PhysRevB.101.195147,bessho2020topological,okuma2020nonhermitian,PhysRevB.102.205118,okuma2020quantum,kawabata2020topological,leeAnatomySkinModes2019b,PhysRevLett.123.016805,PhysRevResearch.2.023265,PhysRevLett.124.250402,PhysRevB.102.085151,Li:2020aa,Lee:2020aa,lee2020manybody,li2020impurity,arouca2020unconventional,lee2020exceptional,pan2020pointgap,PhysRevLett.122.237601,PhysRevResearch.1.023013,PhysRevLett.124.066602,longhi2020stochastic,PhysRevB.102.201103,PhysRevB.100.054301,liu2020localization,PhysRevB.102.075404,xiongWhyDoesBulk2018c,PhysRevB.97.121401,dengNonBlochTopologicalInvariants2019a,PhysRevB.100.035102,li2020twodimensional,PhysRevB.101.121116,PhysRevB.101.235150,PhysRevLett.125.118001,PhysRevResearch.2.022062,PhysRevA.102.023308,cao2020nonhermitian,claes2020skin,ma2020quantum,okugawa2020secondorder,PhysRevLett.125.123902,PhysRevLett.125.206402,yu2020unsupervised,yoshida2020rdmft} to Hermitian systems, and is dubbed as {\em Hermitian skin effect} in this paper. We emphasize that although the Bosonic Hamiltonian is Hermitian, its elementary excitations are determined by a non-Hermitian matrix~\cite{xiao2009theory,KAWAGUCHI2012253}. In order to understand the quasiparticles in such systems, the so-called generalized Brillouin zone (GBZ) theory~\cite{yaoEdgeStatesTopological2018b,yokomizoNonBlochBandTheory2019a,PhysRevLett.125.126402,PhysRevLett.125.226402} is necessary. We finally provide a symmetry argument for the emergence of Hermitian skin effect. 

{\em Thermodynamic limit.}---We first explain why the traditional thermodynamic limit argument fails in predicting the Hermitian skin effect. As shown in Fig.~\ref{F1} (a), it can be noticed that the boundary condition has been changed in the thermodynamic limit we applied, i.e. from OBC to IBC. Once the OBC eigenstates have some nontrivial localized properties (as can be seen in the following example), the thermodynamic limit shown in Fig.~\ref{F1} (a) can no longer describe the corresponding system.

In addition to the thermodynamic limit ($L=Na\rightarrow\infty$) shown in Fig.~\ref{F1} (a), there is another situation when the lattice length $L$ is fixed, i.e. $a=L/N\rightarrow 0$, as shown in Fig.~\ref{F1} (b)~\cite{PhysRevB.96.195133,PhysRevB.98.245423}. Since the system is finite, this thermodynamic limit can be applied to describe the above-mentioned localized properties by using the concept of non-Bloch waves, which are characterized by complex crystal momentum, whose imaginary part represents the localization length and directions~\cite{yaoEdgeStatesTopological2018b}. It has been shown that the asymptotic behavior of OBC Hamiltonians in the second thermodynamic limit can be described by the GBZ theory~\cite{yaoEdgeStatesTopological2018b,yokomizoNonBlochBandTheory2019a,PhysRevLett.125.126402,PhysRevLett.125.226402}, as shown in Fig.~\ref{F1} (b). In the GBZ theory, the concept of the dispersion relation is still preserved but has been extended, in which both the momentum and the energy can be complex numbers. Note that these non-Bloch waves described by the GBZ theory have no relation to the topological protected boundary states appearing in the topological band theory (see the model below).

{\em Model.---}We consider a one-dimensional (1D) system of coupled bosonic modes, whose particle number is not conserved~\cite{RevModPhys.91.015006}. However, we emphasize that the phenomenon illustrated here applies also for higher-dimensional systems. Under the PBC, the Bloch Hamiltonian proposed in this work reads $\hat{H}_B=C+\frac{1}{2}\sum_{k\in{\rm BZ}}\hat{\Psi}_k^\dag\mathcal{H}_B(k)\hat{\Psi}_k$, where $C=-{\rm Tr}[\mathcal{H}_0(k)]/2$,  $\hat{\Psi}_k=(a_k,b_k,a_{-k}^\dag,b_{-k}^\dag)^t$ and 
\begin{equation}
\mathcal{H}_B(k)=\left(\begin{array}{cc}
\mathcal{H}_0(k) & \Delta(k) \\
\Delta^*(-k) & \mathcal{H}^*_0(-k)
\end{array}\right).
\label{E1}
\end{equation}
Here $\mathcal{H}_0(k)=(t_1+t_2\cos k)\sigma_x+t_2\sin k\sigma_y+2\lambda\sin k\sigma_z$ and $\Delta(k)=\Delta_0\sigma_0$ are $2\times2$ matrices. $a_k$ and $b_k$ annihilate the Bosons $A$ and $B$ with momentum $k$. The Pauli matrix $\sigma_\mu$ acts on the $A/B$ degrees of freedom. In general, $\mathcal{H}_0(k)$ can also have $\mu_0\sigma_0+\mu_z\sigma_z$ term, whose roles will be discussed in the final section. In the current discussion, we assume $\mu_0=\mu_z=0$. When $\Delta_0=0$, the Hamiltonian describes a time-reversal symmetry breaking bosonic Su-Schrieﬀer-Heeger (SSH) model~\cite{PhysRevLett.125.186802}. When $\Delta_0\neq0$, the existence of pairing term implies the particle number is not conserved. Physically, the pairing term can be experimentally realized in magnon~\cite{PhysRevB.89.054420},  superfluid~\cite{KAWAGUCHI2012253,PhysRevA.88.063631,PhysRevA.101.013625,PhysRevLett.115.245302,PhysRevLett.117.045302}, optical~\cite{RevModPhys.91.015006,PhysRevX.6.041026}, and coupled oscillators~\cite{PhysRevX.8.041031}. 

Although the bosonic Hamiltonian is Hermitian, its dynamical properties (or quasiparticles) are determined by a non-Hermitian matrix due to the bosonic commutation relation~\cite{xiao2009theory,KAWAGUCHI2012253}. An intuitive way to understand this fact is to consider the Heisenberg equation of the field $\hat{\Psi}_k=(a_k,b_k,a_{-k}^\dag,b_{-k}^\dag)^t$, i.e.
\begin{equation}
i\frac{d}{dt}\hat{\Psi}_k(t)=\mathcal{M}_B(k)\hat{\Psi}_k(t),\quad \mathcal{M}_B(k)=I_k\mathcal{H}_B(k), 
\label{E2}
\end{equation}
where $I_k=\diag(1,1,-1,-1)$ and the explite from of $\mathcal{M}_B(k)$ reads $\mathcal{M}_B(k)=[(t_1+t_2\cos k)\sigma_x+t_2\sin k\sigma_y]\tau_z+2\lambda\sin k\sigma_z+i\Delta_0\tau_y$, where the Pauli matrix $\tau_\mu$ describes the particle-hole degrees of freedom. The eigenvalues of $\mathcal{M}_B(k)$ are $\pm E_{+}(k)$ and $\pm E_-(k)$ with 
\begin{equation}
E_{\pm}(k)=\sqrt{t_1^2+t_2^2+2t_1t_2\cos k+(2\lambda\sin k\pm i\Delta_0)^2}.
\label{E4}
\end{equation}
Since $\mathcal{M}_B(k)$ is non-Hermitian, the corresponding eigenvalues in Eq.~\ref{E4} can be complex. If the imaginary part of $E_\pm(k)$ is nonzero, it will indicate the emergence of dynamical instability~\cite{PhysRevA.64.061603,KAWAGUCHI2012253}, since the excitations will decay or amplify with time. 

As shown in Fig.~\ref{F2}~(a1), we plot the dynamical instability region of $\mathcal{M}_B(k)$ with PBC and with $t_2=1, \Delta_0=1/3$. The strength of the color represents the largest imaginary part of the eigenvalues of $\mathcal{M}_B(k)$ for $k\in[-\pi,\pi]$. Only when $\lambda=0$ (which is represented by the dashed black line), the dynamical instability disappears. We also plot several examples of the PBC energy spectrum in Fig.~\ref{F2}~(b)-(d) with gray points. Here $t_2$ and $\Delta_0$ are chosen the same as Fig.~\ref{F2} (a), and the other parameters are shown the ones on the above of these sub-figures. One can notice that as the increasing of $\lambda$, the largest imaginary part of $E_{PBC}$ also increases. However, when we calculate the spectrum of $\mathcal{M}_B(k)$ with OBC, some of the results are distinct from the ones with PBC, which are shown in Fig.~\ref{F2}~(b)-(d) with black points. Especially, when $\lambda=1/2$, all the OBC energy spectrum becomes real, which indicates the disappearance of dynamical instability although the PBC result predicts the emergence of dynamical instability. Actually, as shown in Fig.~\ref{F2}~(a2), the OBC result has a large dynamically stable region (white region), in which all the OBC eigenvalues have no imaginary part. Both the differences of the dynamical instability regions and the corresponding distinct energy spectrum between PBC and OBC imply the Bloch's theorem may be non-perturbatively broken down in such a system under certain parameter regions ($\lambda\neq0$). 

\begin{figure}[t]
	\centerline{\includegraphics[width=1\linewidth]{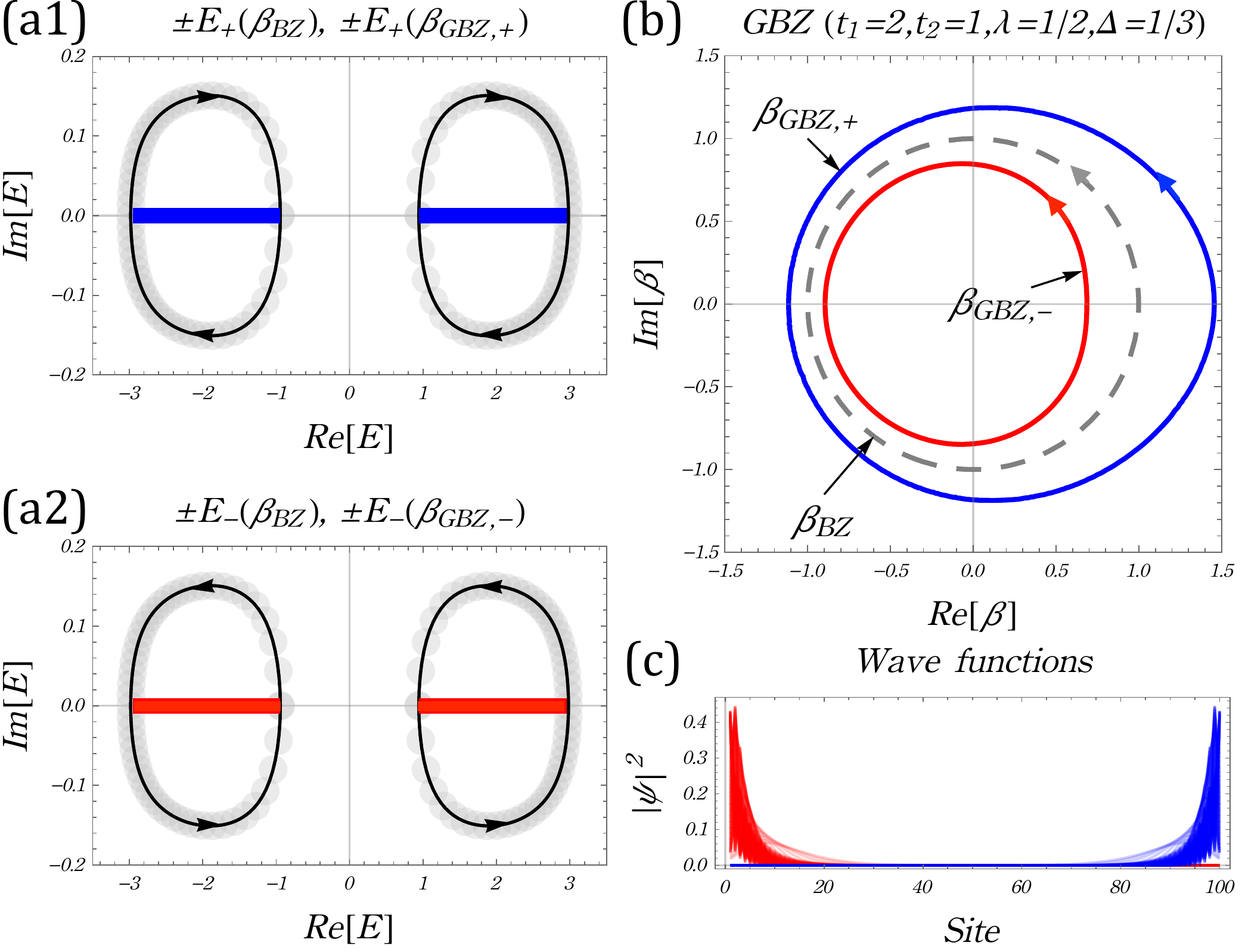}}
	\caption{(a) shows the BZ and GBZ spectrum of $\mathcal{M}_\pm(k)$ in Eq.~\ref{E6}. (b) shows the corresponding BZ and GBZ. (c) shows all the eigenstates of $\mathcal{M_\pm}(k)$ with OBC (blue/red colors). The parameters are shown in (b). 
		\label{F3}}
\end{figure}

{\em The GBZ theory.}---Now we use the GBZ theory to explain the above spectrum behavior~\cite{yaoEdgeStatesTopological2018b,yokomizoNonBlochBandTheory2019a,PhysRevLett.125.126402,PhysRevLett.125.226402}. By applying the following unitary transformation, $\mathcal{M}_B(k)$ can be decoupled into the following two independent blocks, i.e.
\begin{equation}
	U_k^{-1}\mathcal{M}_{B}(k)U_k=\left(\begin{array}{cc}
		\mathcal{M}_{+}(k) & 0 \\
		0 & \mathcal{M}_{-}(k), 
	\end{array}\right), 
	\label{E5}
\end{equation}
where 
\begin{equation}\begin{aligned}
		\mathcal{M}_{\pm}(k)=&-(t_1+t_2\cos k)\sigma_x-t_2\sin k\sigma_y\\
		&+(2\lambda\sin k\pm i\Delta)\sigma_z,\label{E6}
\end{aligned}\end{equation}
and  
\begin{equation}
	U_k=\frac{1}{\sqrt{2}}\left(\begin{array}{cc}
		-i\sigma_z & i\sigma_z \\
		1 & 1, 
	\end{array}\right).
\end{equation}
The eigenvalues of $\mathcal{M}_+(k)$ and $\mathcal{M}_-(k)$ are $\pm E_{+}(k)$ and $\pm E_{-}(k)$, respectively. In order to understand the spectrum difference, we extend the momentum $k$ from $\mathbb{R}$ to $\mathbb{C}$. Since $E_\pm(k)$ are periodic functions, it is convenient to use the new variable $\beta=e^{ik}$~\cite{yaoEdgeStatesTopological2018b}. When $\beta$ is extended from the unit circle (or BZ which is labeled by $\beta_{BZ}=e^{ik}$ with $k\in\mathbb{R}$) to the entire complex plane, the Bloch band is also extended to the non-Bloch band,  $\pm E_{\pm }(\beta=e^{ik}\in \mathbb{C})$. As shown in Fig.~\ref{F3} (a1) and (a2), $\pm E_{\pm}(\beta_{BZ})$ are plotted with solid black lines, whose arrows represent the orientation of the spectrum path when $k$ evolves from $-\pi$ to $\pi$. As a comparison, we also plot the numerical PBC spectrum with gray points. Since the OBC spectrum is distinct from the PBC spectrum, a simple explanation is that the possible values of $\beta$ are changed from $\beta_{BZ}$ to other loops on the complex plane. 

The above augment motivates the GBZ theory. As shown in Fig.~\ref{F3} (b), the dashed gray line represents the BZ in the complex plane, and any state with momentum laying on it corresponds to the Bloch waves, or extended states. The blue/red lines represent the GBZs for the $\mathcal{M}_{\pm}(k)$ with OBC. Since the red loop is inside the unit circle (BZ), all its corresponding eigenstates are localized at the left boundary, i.e. $x=0$. And the blue loop, which is outside the unit circle, corresponds to the right localized states. This is consistent with the numerical diagonalization results shown in Fig.~\ref{F3} (c), where the blue/red colors represent {\em all} the eigenstates of $\mathcal{M}_{\pm}(k)$ with OBC, respectively. Having the GBZ, the asymptotic spectrum can also be obtained by putting $\beta_{GBZ,+}$/$\beta_{GBZ,-}$ into $\pm E_{+}(\beta)$/$\pm E_{-}(\beta)$, which are shown in Fig.~\ref{F3} (a1/a2) with blue/red lines. From the results shown in (a1) and (a2), one can find that the energy spectrum between $\pm E_+(\beta_{GBZ,+})$ and $\pm E_-(\beta_{GBZ,-})$ are degenerate. 

{\em Elementary excitations.}---In order to further illustrate the non-perturbative breakdown of Bloch's theorem, we proceed to the discussion of elementary excitations. We use $M_B$ to label the OBC Hamiltonian of $\mathcal{M}_B(k)$ in the particle-hole basis, whose dimension is $4N\times 4N$, where $N$ represents the lattice size. If the OBC spectrum is real and has no degeneracy, the quasiparticle excitations of the Bosonic Hamiltonian are the eigenstates of $M_B$ with the following normalization condition~\cite{xiao2009theory,KAWAGUCHI2012253}, 
\begin{equation}
	\langle \Phi_n|I_r|\Phi_n\rangle=1, \quad \langle \bar{\Phi}_{\bar{n}}|I_r|\bar{\Phi}_{\bar{n}}\rangle=-1, 
\end{equation}
where $I_r=\tau_z\otimes\mathbb{1}_{2N}$. Here $|\Phi_n\rangle$ is the particle excitation with energy $E_n$, and $|\bar{\Phi}_{\bar{n}}\rangle=\mathcal{\bar{C}}|\Phi_n\rangle$ is the hole excitation with energy $E_{\bar{n}}=-E_n^*$, where $\mathcal{\bar{C}}=\tau_x\otimes\mathbb{1}_{2N}\mathcal{K}^*$ is the representation of the anomalous particle-hole symmetry (PHS$^\dag$) and $\mathcal{K}^*$ is the complex conjugate operator~\cite{PhysRevX.8.031079,PhysRevX.9.041015,PhysRevB.99.125103,PhysRevB.100.144106,PhysRevB.99.235112,PhysRevLett.123.206404}. One can check that these normalized eigenstates satisfy the Bosonic commutation relation. 

Back to our model, since the eigenstate of $M_B$ has two-fold degeneracy. We first need to find two orthogonal and normalized basis. This can be done by using the real space transformation of Eq.~\ref{E5}. For any eigenstate $|\phi_{n}\rangle$ of $M_B$ with eigenvalue $E_n$, we can project it into the following two eigenstates, 
\begin{equation}
	|\phi_{n,\pm}\rangle=N_{\pm}U_rP_\pm U_r^{-1}	|\phi_{n}\rangle, 
\end{equation}  
where 
\begin{equation}
	U_r=\frac{1}{\sqrt{2}}\left(\begin{array}{cc}
		-i\mathbb{1}_N\otimes\sigma_z & i\mathbb{1}_N\otimes\sigma_z \\
		\mathbb{1}_{2N} & \mathbb{1}_{2N} 
	\end{array}\right),
\end{equation}
and $N_\pm$ is the normalization factor satisfying $N_\pm^2=\langle\phi_{n,\pm}|\phi_{n,\pm}\rangle$, and 
\begin{equation}
P_{+}=\left(\begin{array}{cc}
	1 & 0 \\
	0 & 0
\end{array}\right)\otimes  \mathbb{1}_{2N},\quad  P_{-}=\left(\begin{array}{cc}
0 & 0 \\
0 & 1 
\end{array}\right)\otimes  \mathbb{1}_{2N}. 
\end{equation}
Under this projection, $|\phi_{n,\pm}\rangle$ are the superposition of the non-Bloch waves on the $\beta_{GBZ,\pm}$. On the other hand, from  $U_r^{-1}I_rU_r=-\tau_x\otimes \mathbb{1}_{2N}$, we have  $\langle\phi_{n,+}|I_r|\phi_{n,+}\rangle=\langle\phi_{n,-}|I_r|\phi_{n,-}\rangle=0$. Therefore solving the following matrix 
\begin{equation}
	I_{r,n}=\left(\begin{array}{cc}
	0 & \lambda_ne^{i\theta_n} \\
	\lambda_ne^{-i\theta_n}  & 0
\end{array}\right), \lambda_ne^{i\theta_n}:=\langle\phi_{n,+}|I_r|\phi_{n,-}\rangle,
\end{equation}
we can obtain the corresponding two Bosonic excitations with energy  $E_n$, 
\begin{equation}\begin{aligned}
	&|\Phi_{n,+}\rangle=\frac{1}{\sqrt{2\lambda_n}}(e^{i\theta_n}|\phi_{n,+}\rangle+|\phi_{n,-}\rangle),\\
	&|\bar{\Phi}_{\bar{n},-}\rangle=\frac{1}{\sqrt{2\lambda_n}}(e^{i\theta_n}|\phi_{n,+}\rangle-|\phi_{n,-}\rangle).
\end{aligned}\end{equation}
Here $|\Phi_{n,+}/\bar{\Phi}_{\bar{n},-}\rangle$ represent the particle/hole excitations. It can be verified that all the eigenstates obtained from the above method satisfy the Bosonic commutation relation, i.e.   
\begin{equation}\begin{aligned}
&\langle \Phi_{n,\sigma}|I_r|\Phi_{m,\rho}\rangle=\delta_{mn}\delta_{\sigma\rho}, \quad \langle \bar{\Phi}_{\bar{n},\sigma}|I_r|\bar{\Phi}_{\bar{m},\rho}\rangle=-\delta_{\bar{m}\bar{n}}\delta_{\sigma\rho},\\
&\langle\Phi_{n,\sigma}|I_r|\bar{\Phi}_{\bar{m},\rho}\rangle=0, \qquad~~~~~ \langle \bar{\Phi}_{\bar{n},\sigma}|I_r|\Phi_{m,\rho}\rangle=0.
\end{aligned}\end{equation}
From the above results, one can see that the wave function of the elementary excitation is a equal superposition of two non-Bloch waves with the same localization length but opposite directions. As a result, the excitations are localized at both boundaries with equal weight. 

{\em Symmetries and Hermitian $Z_2$ skin effect.}---Now we analyze why Bloch's theorem can be non-perturbatively broken down in our model. It turns out that the inversion symmetry (IS) $\mathcal{P}$ and PHS$^\dag$ $\mathcal{\bar{C}}_+=\tau_x\mathcal{K}^*$ play a crucial role here, where the lower index represents the following constraint  $U_\mathcal{\bar{C}_+}U_\mathcal{\bar{C}_+}^*=+1$. In our example, the IS has the following two different types of representations, i.e. 
\begin{equation}
	\mathcal{P}_1=\sigma_y\tau_y, \qquad \mathcal{P}_2=\sigma_x,
\end{equation} 
Here we first discuss the role of $\mathcal{P}_1$ representation. According to their combination  $(\mathcal{P}_1\mathcal{\bar{C}})_-=\sigma_y\tau_z\mathcal{K}^*$, we know that this symmetry provides an additional band index to each eigenstate, which can be labeled by $\uparrow_{(\mathcal{P}_1\mathcal{\bar{C}})_-}$ and $\downarrow_{(\mathcal{P}_1\mathcal{\bar{C}})_-}$. Therefore, if $|E,\beta,\uparrow_{(\mathcal{P}_1\mathcal{\bar{C}})_-}\rangle$ is an eigenstate of the non-Bloch Hamiltonian with energy $E$ and localization length $|\beta|$, IS maps it to $|E,1/\beta,\downarrow_{(\mathcal{P}_1\mathcal{\bar{C}})_-}\rangle$. This imposes the following GBZ condition~\cite{PhysRevLett.125.186802}
\begin{equation}
	|\beta_{p-1}|=|\beta_p|,\quad |\beta_{p+1}|=|\beta_{p+2}|, 
\end{equation}
where $\beta_{i}$ is the $i$the largest root of $\det[E-\mathcal{M}_B(\beta)]=0$ ordered by the absolute value, and $p$ is the order of the pole of $\det[E-\mathcal{M}_B(\beta)]=0$. Previous works have shown that the above GBZ condition wil induce a $Z_2$ skin effect~\cite{okumaTopologicalOriginNonHermitian2019,PhysRevLett.125.186802,PhysRevB.101.195147}. 

Now, we will use the same procedure to discuss the effect of the following term $\mu_0\tau_z+\mu_z\sigma_z\tau_z$. (i) When $\mu_0=0,\mu_z\neq0$, the representation of $\mathcal{P}_1$ in unchanged. Therefore, the system also has the $Z_2$ skin effect. However, a numerical calculation shows that the dynamically stable region will disappear under the OBC. (ii) When $\mu_0\neq0,\mu_z=0$, the corresponding representation becomes $\mathcal{P}_2$, which is commutative to the PHS$^\dag$, e.g. $[\mathcal{\bar{C}}_+,\mathcal{P}_2]=0$. This means the inversion symmetry connects the same band with the same energy. This will trivialize the $Z_2$ skin effect. (iii) When $\mu_0\neq0,\mu_z\neq 0$, the IS is broken. In this case, the system has a $Z$ skin effect. We note that from $Z_2$ to $Z$ or $0$, critical skin appears~\cite{Li:2020aa,PhysRevLett.123.097701,PhysRevLett.125.226402,PhysRevResearch.2.043167}. 

Finally, we will provide a sufficient condition for the emergence of skin effects in the dynamically stable system. From the pseudo-Hermiticity of $M_B$, i.e.  $I_r^{-1}M_BI_r=M^\dag_B$, we know that if $|\Phi_n\rangle$ is a right-eigenstate of $M_B$ with energy $E_n\in\mathbb{R}$, then $I_r|\Phi_n\rangle$ must be a left-eigenstate~\cite{Brody_2013} of $M_B$, i.e. $M_B^\dag(I_r^{-1}|\Phi_n\rangle)=E_n^*(I_r^{-1}|\Phi_n\rangle)$. Previous studies have shown that if a right-eigenstate is localized at one boundary, the corresponding left-eigenstate must be localized at the opposite direction~\cite{PhysRevLett.125.186802}. However, the diagonal matrix $I_r^{-1}$ cannot change the localization properties. As a result, in order to emerge the skin effect, there must exist another degenerate state, whose localization is opposite to $|\Phi_n\rangle$. This is the Hermitian $Z_2$ skin effect discussed in our example, where the quasiparticles are the superposition of these non-Bloch waves and are localized at the two boundaries with equal weight. 

{\em Discussions and conclusions.}---In summary, by using a concrete model, our work reveals a new paradigm in condensed matter physics, that is, all quasiparticles in the Hermitian bosonic system can be non-Bloch waves, which are not extended but localized at two distinct boundaries. In such a system the traditional Bloch's theorem, which plays a fundamental role in condensed matter physics, is non-perturbatively broken down. Since these quasiparticles can be dynamically stable, how to understand the corresponding physical response is the next step of further research. 

{\em Acknowledgments.}---The author thanks the valuable discussion with Chen Fang and Kai Zhang. 

{\em Note added.}---Recently, we became aware of a related work~\cite{yokomizo2020nonbloch}.

\bibliography{aGBZ}

\begin{thebibliography}{97}%
\makeatletter
\providecommand \@ifxundefined [1]{%
 \@ifx{#1\undefined}
}%
\providecommand \@ifnum [1]{%
 \ifnum #1\expandafter \@firstoftwo
 \else \expandafter \@secondoftwo
 \fi
}%
\providecommand \@ifx [1]{%
 \ifx #1\expandafter \@firstoftwo
 \else \expandafter \@secondoftwo
 \fi
}%
\providecommand \natexlab [1]{#1}%
\providecommand \enquote  [1]{``#1''}%
\providecommand \bibnamefont  [1]{#1}%
\providecommand \bibfnamefont [1]{#1}%
\providecommand \citenamefont [1]{#1}%
\providecommand \href@noop [0]{\@secondoftwo}%
\providecommand \href [0]{\begingroup \@sanitize@url \@href}%
\providecommand \@href[1]{\@@startlink{#1}\@@href}%
\providecommand \@@href[1]{\endgroup#1\@@endlink}%
\providecommand \@sanitize@url [0]{\catcode `\\12\catcode `\$12\catcode
  `\&12\catcode `\#12\catcode `\^12\catcode `\_12\catcode `\%12\relax}%
\providecommand \@@startlink[1]{}%
\providecommand \@@endlink[0]{}%
\providecommand \url  [0]{\begingroup\@sanitize@url \@url }%
\providecommand \@url [1]{\endgroup\@href {#1}{\urlprefix }}%
\providecommand \urlprefix  [0]{URL }%
\providecommand \Eprint [0]{\href }%
\providecommand \doibase [0]{http://dx.doi.org/}%
\providecommand \selectlanguage [0]{\@gobble}%
\providecommand \bibinfo  [0]{\@secondoftwo}%
\providecommand \bibfield  [0]{\@secondoftwo}%
\providecommand \translation [1]{[#1]}%
\providecommand \BibitemOpen [0]{}%
\providecommand \bibitemStop [0]{}%
\providecommand \bibitemNoStop [0]{.\EOS\space}%
\providecommand \EOS [0]{\spacefactor3000\relax}%
\providecommand \BibitemShut  [1]{\csname bibitem#1\endcsname}%
\let\auto@bib@innerbib\@empty
\bibitem [{\citenamefont {Bloch}(1929)}]{Bloch_theorem}%
  \BibitemOpen
  \bibfield  {author} {\bibinfo {author} {\bibfnamefont {F.}~\bibnamefont
  {Bloch}},\ }\href@noop {} {\bibfield  {journal} {\bibinfo  {journal}
  {Zeitschrift f{\"u}r Physik}\ }\textbf {\bibinfo {volume} {55}},\ \bibinfo
  {pages} {555} (\bibinfo {year} {1929})}\BibitemShut {NoStop}%
\bibitem [{\citenamefont {Kittel}(1976)}]{textbook1}%
  \BibitemOpen
  \bibfield  {author} {\bibinfo {author} {\bibfnamefont {C.}~\bibnamefont
  {Kittel}},\ }\href@noop {} {\emph {\bibinfo {title} {Introduction to solid
  state physics. Fifth edition}}}\ (\bibinfo {year} {1976})\BibitemShut
  {NoStop}%
\bibitem [{\citenamefont {Ashcroft}\ and\ \citenamefont
  {Mermin}(1976)}]{Ashcroft76}%
  \BibitemOpen
  \bibfield  {author} {\bibinfo {author} {\bibfnamefont {N.~W.}\ \bibnamefont
  {Ashcroft}}\ and\ \bibinfo {author} {\bibfnamefont {N.~D.}\ \bibnamefont
  {Mermin}},\ }\href@noop {} {\emph {\bibinfo {title} {{S}olid {S}tate
  {P}hysics}}}\ (\bibinfo  {publisher} {Holt-Saunders},\ \bibinfo {year}
  {1976})\BibitemShut {NoStop}%
\bibitem [{\citenamefont {Coleman}(2015)}]{coleman_2015}%
  \BibitemOpen
  \bibfield  {author} {\bibinfo {author} {\bibfnamefont {P.}~\bibnamefont
  {Coleman}},\ }\href {\doibase 10.1017/CBO9781139020916} {\emph {\bibinfo
  {title} {Introduction to Many-Body Physics}}}\ (\bibinfo  {publisher}
  {Cambridge University Press},\ \bibinfo {year} {2015})\BibitemShut {NoStop}%
\bibitem [{\citenamefont {Nozieres}\ and\ \citenamefont
  {Pines}(1999)}]{Pines_1999}%
  \BibitemOpen
  \bibfield  {author} {\bibinfo {author} {\bibfnamefont {P.}~\bibnamefont
  {Nozieres}}\ and\ \bibinfo {author} {\bibfnamefont {D.}~\bibnamefont
  {Pines}},\ }\href@noop {} {\emph {\bibinfo {title} {Theory of quantum
  liquids}}}\ (\bibinfo  {publisher} {Cambridge University Press},\ \bibinfo
  {year} {1999})\BibitemShut {NoStop}%
\bibitem [{\citenamefont {Bardeen}\ \emph {et~al.}(1957)\citenamefont
  {Bardeen}, \citenamefont {Cooper},\ and\ \citenamefont
  {Schrieffer}}]{PhysRev.108.1175}%
  \BibitemOpen
  \bibfield  {author} {\bibinfo {author} {\bibfnamefont {J.}~\bibnamefont
  {Bardeen}}, \bibinfo {author} {\bibfnamefont {L.~N.}\ \bibnamefont {Cooper}},
  \ and\ \bibinfo {author} {\bibfnamefont {J.~R.}\ \bibnamefont {Schrieffer}},\
  }\href {\doibase 10.1103/PhysRev.108.1175} {\bibfield  {journal} {\bibinfo
  {journal} {Phys. Rev.}\ }\textbf {\bibinfo {volume} {108}},\ \bibinfo {pages}
  {1175} (\bibinfo {year} {1957})}\BibitemShut {NoStop}%
\bibitem [{Note1()}]{Note1}%
  \BibitemOpen
  \bibinfo {note} {Means the wavefunction must be bounded at the infinity $\pm
  \infty $}\BibitemShut {NoStop}%
\bibitem [{\citenamefont {Mahan}(2000)}]{mah00}%
  \BibitemOpen
  \bibfield  {author} {\bibinfo {author} {\bibfnamefont {G.~D.}\ \bibnamefont
  {Mahan}},\ }\href@noop {} {\emph {\bibinfo {title} {Many Particle Physics,
  Third Edition}}}\ (\bibinfo  {publisher} {Plenum},\ \bibinfo {address} {New
  York},\ \bibinfo {year} {2000})\BibitemShut {NoStop}%
\bibitem [{\citenamefont {Altland}\ and\ \citenamefont
  {Simons}(2010)}]{altland_simons_2010}%
  \BibitemOpen
  \bibfield  {author} {\bibinfo {author} {\bibfnamefont {A.}~\bibnamefont
  {Altland}}\ and\ \bibinfo {author} {\bibfnamefont {B.~D.}\ \bibnamefont
  {Simons}},\ }\href {\doibase 10.1017/CBO9780511789984} {\emph {\bibinfo
  {title} {Condensed Matter Field Theory}}},\ \bibinfo {edition} {2nd}\ ed.\
  (\bibinfo  {publisher} {Cambridge University Press},\ \bibinfo {year}
  {2010})\BibitemShut {NoStop}%
\bibitem [{\citenamefont {Bernevig}\ and\ \citenamefont
  {Hughes}(2013)}]{10.2307/j.ctt19cc2gc}%
  \BibitemOpen
  \bibfield  {author} {\bibinfo {author} {\bibfnamefont {B.~A.}\ \bibnamefont
  {Bernevig}}\ and\ \bibinfo {author} {\bibfnamefont {T.~L.}\ \bibnamefont
  {Hughes}},\ }\href {http://www.jstor.org/stable/j.ctt19cc2gc} {\emph
  {\bibinfo {title} {Topological Insulators and Topological
  Superconductors}}},\ \bibinfo {edition} {stu - student edition}\ ed.\
  (\bibinfo  {publisher} {Princeton University Press},\ \bibinfo {year}
  {2013})\BibitemShut {NoStop}%
\bibitem [{\citenamefont {Born}\ and\ \citenamefont
  {Huang}(1988)}]{born1988dynamical}%
  \BibitemOpen
  \bibfield  {author} {\bibinfo {author} {\bibfnamefont {M.}~\bibnamefont
  {Born}}\ and\ \bibinfo {author} {\bibfnamefont {K.}~\bibnamefont {Huang}},\
  }\href {https://books.google.co.kr/books?id=5q9iRttaaDAC} {\emph {\bibinfo
  {title} {Dynamical Theory of Crystal Lattices}}},\ International series of
  monographs on physics\ (\bibinfo  {publisher} {Clarendon Press},\ \bibinfo
  {year} {1988})\BibitemShut {NoStop}%
\bibitem [{\citenamefont {Anderson}(1997)}]{anderson1997concepts}%
  \BibitemOpen
  \bibfield  {author} {\bibinfo {author} {\bibfnamefont {P.}~\bibnamefont
  {Anderson}},\ }\href {https://books.google.co.kr/books?id=2NEKzzbMbRgC}
  {\emph {\bibinfo {title} {Concepts in Solids: Lectures on the Theory of
  Solids}}},\ Advanced book classics series\ (\bibinfo  {publisher} {World
  Scientific},\ \bibinfo {year} {1997})\BibitemShut {NoStop}%
\bibitem [{\citenamefont {Zhai}(2021)}]{zhai2021ultracold}%
  \BibitemOpen
  \bibfield  {author} {\bibinfo {author} {\bibfnamefont {H.}~\bibnamefont
  {Zhai}},\ }\href {https://books.google.co.kr/books?id=IZvWzQEACAAJ} {\emph
  {\bibinfo {title} {Ultracold Atomic Physics}}}\ (\bibinfo  {publisher}
  {Cambridge University Press},\ \bibinfo {year} {2021})\BibitemShut {NoStop}%
\bibitem [{\citenamefont {Yao}\ and\ \citenamefont
  {Wang}(2018)}]{yaoEdgeStatesTopological2018b}%
  \BibitemOpen
  \bibfield  {author} {\bibinfo {author} {\bibfnamefont {S.}~\bibnamefont
  {Yao}}\ and\ \bibinfo {author} {\bibfnamefont {Z.}~\bibnamefont {Wang}},\
  }\href {\doibase 10.1103/PhysRevLett.121.086803} {\bibfield  {journal}
  {\bibinfo  {journal} {Phys. Rev. Lett.}\ }\textbf {\bibinfo {volume} {121}},\
  \bibinfo {pages} {086803} (\bibinfo {year} {2018})}\BibitemShut {NoStop}%
\bibitem [{\citenamefont {Yao}\ \emph {et~al.}(2018)\citenamefont {Yao},
  \citenamefont {Song},\ and\ \citenamefont
  {Wang}}]{yaoNonHermitianChernBands2018b}%
  \BibitemOpen
  \bibfield  {author} {\bibinfo {author} {\bibfnamefont {S.}~\bibnamefont
  {Yao}}, \bibinfo {author} {\bibfnamefont {F.}~\bibnamefont {Song}}, \ and\
  \bibinfo {author} {\bibfnamefont {Z.}~\bibnamefont {Wang}},\ }\href {\doibase
  10.1103/PhysRevLett.121.136802} {\bibfield  {journal} {\bibinfo  {journal}
  {Phys. Rev. Lett.}\ }\textbf {\bibinfo {volume} {121}},\ \bibinfo {pages}
  {136802} (\bibinfo {year} {2018})}\BibitemShut {NoStop}%
\bibitem [{\citenamefont {Song}\ \emph
  {et~al.}(2019{\natexlab{a}})\citenamefont {Song}, \citenamefont {Yao},\ and\
  \citenamefont {Wang}}]{PhysRevLett.123.170401}%
  \BibitemOpen
  \bibfield  {author} {\bibinfo {author} {\bibfnamefont {F.}~\bibnamefont
  {Song}}, \bibinfo {author} {\bibfnamefont {S.}~\bibnamefont {Yao}}, \ and\
  \bibinfo {author} {\bibfnamefont {Z.}~\bibnamefont {Wang}},\ }\href {\doibase
  10.1103/PhysRevLett.123.170401} {\bibfield  {journal} {\bibinfo  {journal}
  {Phys. Rev. Lett.}\ }\textbf {\bibinfo {volume} {123}},\ \bibinfo {pages}
  {170401} (\bibinfo {year} {2019}{\natexlab{a}})}\BibitemShut {NoStop}%
\bibitem [{\citenamefont {Song}\ \emph
  {et~al.}(2019{\natexlab{b}})\citenamefont {Song}, \citenamefont {Yao},\ and\
  \citenamefont {Wang}}]{PhysRevLett.123.246801}%
  \BibitemOpen
  \bibfield  {author} {\bibinfo {author} {\bibfnamefont {F.}~\bibnamefont
  {Song}}, \bibinfo {author} {\bibfnamefont {S.}~\bibnamefont {Yao}}, \ and\
  \bibinfo {author} {\bibfnamefont {Z.}~\bibnamefont {Wang}},\ }\href {\doibase
  10.1103/PhysRevLett.123.246801} {\bibfield  {journal} {\bibinfo  {journal}
  {Phys. Rev. Lett.}\ }\textbf {\bibinfo {volume} {123}},\ \bibinfo {pages}
  {246801} (\bibinfo {year} {2019}{\natexlab{b}})}\BibitemShut {NoStop}%
\bibitem [{\citenamefont {Xue}\ \emph {et~al.}(2020)\citenamefont {Xue},
  \citenamefont {Li}, \citenamefont {Hu}, \citenamefont {Song},\ and\
  \citenamefont {Wang}}]{xue2020nonhermitian}%
  \BibitemOpen
  \bibfield  {author} {\bibinfo {author} {\bibfnamefont {W.-T.}\ \bibnamefont
  {Xue}}, \bibinfo {author} {\bibfnamefont {M.-R.}\ \bibnamefont {Li}},
  \bibinfo {author} {\bibfnamefont {Y.-M.}\ \bibnamefont {Hu}}, \bibinfo
  {author} {\bibfnamefont {F.}~\bibnamefont {Song}}, \ and\ \bibinfo {author}
  {\bibfnamefont {Z.}~\bibnamefont {Wang}},\ }\href@noop {} {\  (\bibinfo
  {year} {2020})},\ \Eprint {http://arxiv.org/abs/2004.09529}
  {arXiv:2004.09529} \BibitemShut {NoStop}%
\bibitem [{\citenamefont {Xiao}\ \emph {et~al.}(2020)\citenamefont {Xiao},
  \citenamefont {Deng}, \citenamefont {Wang}, \citenamefont {Zhu},
  \citenamefont {Wang}, \citenamefont {Yi},\ and\ \citenamefont
  {Xue}}]{Xiao:2020aa}%
  \BibitemOpen
  \bibfield  {author} {\bibinfo {author} {\bibfnamefont {L.}~\bibnamefont
  {Xiao}}, \bibinfo {author} {\bibfnamefont {T.}~\bibnamefont {Deng}}, \bibinfo
  {author} {\bibfnamefont {K.}~\bibnamefont {Wang}}, \bibinfo {author}
  {\bibfnamefont {G.}~\bibnamefont {Zhu}}, \bibinfo {author} {\bibfnamefont
  {Z.}~\bibnamefont {Wang}}, \bibinfo {author} {\bibfnamefont {W.}~\bibnamefont
  {Yi}}, \ and\ \bibinfo {author} {\bibfnamefont {P.}~\bibnamefont {Xue}},\
  }\href {\doibase 10.1038/s41567-020-0836-6} {\bibfield  {journal} {\bibinfo
  {journal} {Nature Physics}\ }\textbf {\bibinfo {volume} {16}},\ \bibinfo
  {pages} {761} (\bibinfo {year} {2020})}\BibitemShut {NoStop}%
\bibitem [{\citenamefont {Yokomizo}\ and\ \citenamefont
  {Murakami}(2019)}]{yokomizoNonBlochBandTheory2019a}%
  \BibitemOpen
  \bibfield  {author} {\bibinfo {author} {\bibfnamefont {K.}~\bibnamefont
  {Yokomizo}}\ and\ \bibinfo {author} {\bibfnamefont {S.}~\bibnamefont
  {Murakami}},\ }\href {\doibase 10.1103/PhysRevLett.123.066404} {\bibfield
  {journal} {\bibinfo  {journal} {Phys. Rev. Lett.}\ }\textbf {\bibinfo
  {volume} {123}},\ \bibinfo {pages} {066404} (\bibinfo {year}
  {2019})}\BibitemShut {NoStop}%
\bibitem [{\citenamefont {Yokomizo}\ and\ \citenamefont
  {Murakami}(2020{\natexlab{a}})}]{PhysRevResearch.2.043045}%
  \BibitemOpen
  \bibfield  {author} {\bibinfo {author} {\bibfnamefont {K.}~\bibnamefont
  {Yokomizo}}\ and\ \bibinfo {author} {\bibfnamefont {S.}~\bibnamefont
  {Murakami}},\ }\href {\doibase 10.1103/PhysRevResearch.2.043045} {\bibfield
  {journal} {\bibinfo  {journal} {Phys. Rev. Research}\ }\textbf {\bibinfo
  {volume} {2}},\ \bibinfo {pages} {043045} (\bibinfo {year}
  {2020}{\natexlab{a}})}\BibitemShut {NoStop}%
\bibitem [{\citenamefont {Kunst}\ \emph {et~al.}(2018)\citenamefont {Kunst},
  \citenamefont {Edvardsson}, \citenamefont {Budich},\ and\ \citenamefont
  {Bergholtz}}]{PhysRevLett.121.026808}%
  \BibitemOpen
  \bibfield  {author} {\bibinfo {author} {\bibfnamefont {F.~K.}\ \bibnamefont
  {Kunst}}, \bibinfo {author} {\bibfnamefont {E.}~\bibnamefont {Edvardsson}},
  \bibinfo {author} {\bibfnamefont {J.~C.}\ \bibnamefont {Budich}}, \ and\
  \bibinfo {author} {\bibfnamefont {E.~J.}\ \bibnamefont {Bergholtz}},\ }\href
  {\doibase 10.1103/PhysRevLett.121.026808} {\bibfield  {journal} {\bibinfo
  {journal} {Phys. Rev. Lett.}\ }\textbf {\bibinfo {volume} {121}},\ \bibinfo
  {pages} {026808} (\bibinfo {year} {2018})}\BibitemShut {NoStop}%
\bibitem [{\citenamefont {Edvardsson}\ \emph {et~al.}(2020)\citenamefont
  {Edvardsson}, \citenamefont {Kunst}, \citenamefont {Yoshida},\ and\
  \citenamefont {Bergholtz}}]{PhysRevResearch.2.043046}%
  \BibitemOpen
  \bibfield  {author} {\bibinfo {author} {\bibfnamefont {E.}~\bibnamefont
  {Edvardsson}}, \bibinfo {author} {\bibfnamefont {F.~K.}\ \bibnamefont
  {Kunst}}, \bibinfo {author} {\bibfnamefont {T.}~\bibnamefont {Yoshida}}, \
  and\ \bibinfo {author} {\bibfnamefont {E.~J.}\ \bibnamefont {Bergholtz}},\
  }\href {\doibase 10.1103/PhysRevResearch.2.043046} {\bibfield  {journal}
  {\bibinfo  {journal} {Phys. Rev. Research}\ }\textbf {\bibinfo {volume}
  {2}},\ \bibinfo {pages} {043046} (\bibinfo {year} {2020})}\BibitemShut
  {NoStop}%
\bibitem [{\citenamefont {Bergholtz}\ \emph {et~al.}(2020)\citenamefont
  {Bergholtz}, \citenamefont {Budich},\ and\ \citenamefont
  {Kunst}}]{bergholtz2020exceptional}%
  \BibitemOpen
  \bibfield  {author} {\bibinfo {author} {\bibfnamefont {E.~J.}\ \bibnamefont
  {Bergholtz}}, \bibinfo {author} {\bibfnamefont {J.~C.}\ \bibnamefont
  {Budich}}, \ and\ \bibinfo {author} {\bibfnamefont {F.~K.}\ \bibnamefont
  {Kunst}},\ }\href@noop {} {\  (\bibinfo {year} {2020})},\ \Eprint
  {http://arxiv.org/abs/1912.10048} {arXiv:1912.10048} \BibitemShut {NoStop}%
\bibitem [{\citenamefont {Yang}\ \emph
  {et~al.}(2020{\natexlab{a}})\citenamefont {Yang}, \citenamefont {Morampudi},\
  and\ \citenamefont {Bergholtz}}]{yang2020exceptional}%
  \BibitemOpen
  \bibfield  {author} {\bibinfo {author} {\bibfnamefont {K.}~\bibnamefont
  {Yang}}, \bibinfo {author} {\bibfnamefont {S.~C.}\ \bibnamefont {Morampudi}},
  \ and\ \bibinfo {author} {\bibfnamefont {E.~J.}\ \bibnamefont {Bergholtz}},\
  }\href@noop {} {\  (\bibinfo {year} {2020}{\natexlab{a}})},\ \Eprint
  {http://arxiv.org/abs/2007.04329} {arXiv:2007.04329} \BibitemShut {NoStop}%
\bibitem [{\citenamefont {Zhang}\ \emph
  {et~al.}(2020{\natexlab{a}})\citenamefont {Zhang}, \citenamefont {Yang},\
  and\ \citenamefont {Fang}}]{PhysRevLett.125.126402}%
  \BibitemOpen
  \bibfield  {author} {\bibinfo {author} {\bibfnamefont {K.}~\bibnamefont
  {Zhang}}, \bibinfo {author} {\bibfnamefont {Z.}~\bibnamefont {Yang}}, \ and\
  \bibinfo {author} {\bibfnamefont {C.}~\bibnamefont {Fang}},\ }\href {\doibase
  10.1103/PhysRevLett.125.126402} {\bibfield  {journal} {\bibinfo  {journal}
  {Phys. Rev. Lett.}\ }\textbf {\bibinfo {volume} {125}},\ \bibinfo {pages}
  {126402} (\bibinfo {year} {2020}{\natexlab{a}})}\BibitemShut {NoStop}%
\bibitem [{\citenamefont {Yang}\ \emph
  {et~al.}(2020{\natexlab{b}})\citenamefont {Yang}, \citenamefont {Zhang},
  \citenamefont {Fang},\ and\ \citenamefont {Hu}}]{PhysRevLett.125.226402}%
  \BibitemOpen
  \bibfield  {author} {\bibinfo {author} {\bibfnamefont {Z.}~\bibnamefont
  {Yang}}, \bibinfo {author} {\bibfnamefont {K.}~\bibnamefont {Zhang}},
  \bibinfo {author} {\bibfnamefont {C.}~\bibnamefont {Fang}}, \ and\ \bibinfo
  {author} {\bibfnamefont {J.}~\bibnamefont {Hu}},\ }\href {\doibase
  10.1103/PhysRevLett.125.226402} {\bibfield  {journal} {\bibinfo  {journal}
  {Phys. Rev. Lett.}\ }\textbf {\bibinfo {volume} {125}},\ \bibinfo {pages}
  {226402} (\bibinfo {year} {2020}{\natexlab{b}})}\BibitemShut {NoStop}%
\bibitem [{\citenamefont {Yi}\ and\ \citenamefont
  {Yang}(2020)}]{PhysRevLett.125.186802}%
  \BibitemOpen
  \bibfield  {author} {\bibinfo {author} {\bibfnamefont {Y.}~\bibnamefont
  {Yi}}\ and\ \bibinfo {author} {\bibfnamefont {Z.}~\bibnamefont {Yang}},\
  }\href {\doibase 10.1103/PhysRevLett.125.186802} {\bibfield  {journal}
  {\bibinfo  {journal} {Phys. Rev. Lett.}\ }\textbf {\bibinfo {volume} {125}},\
  \bibinfo {pages} {186802} (\bibinfo {year} {2020})}\BibitemShut {NoStop}%
\bibitem [{\citenamefont {Zhang}\ \emph
  {et~al.}(2020{\natexlab{b}})\citenamefont {Zhang}, \citenamefont {Yang},\
  and\ \citenamefont {Hu}}]{PhysRevB.102.045412}%
  \BibitemOpen
  \bibfield  {author} {\bibinfo {author} {\bibfnamefont {Z.}~\bibnamefont
  {Zhang}}, \bibinfo {author} {\bibfnamefont {Z.}~\bibnamefont {Yang}}, \ and\
  \bibinfo {author} {\bibfnamefont {J.}~\bibnamefont {Hu}},\ }\href {\doibase
  10.1103/PhysRevB.102.045412} {\bibfield  {journal} {\bibinfo  {journal}
  {Phys. Rev. B}\ }\textbf {\bibinfo {volume} {102}},\ \bibinfo {pages}
  {045412} (\bibinfo {year} {2020}{\natexlab{b}})}\BibitemShut {NoStop}%
\bibitem [{\citenamefont {Liu}\ \emph {et~al.}(2020{\natexlab{a}})\citenamefont
  {Liu}, \citenamefont {Zhang}, \citenamefont {Yang},\ and\ \citenamefont
  {Chen}}]{PhysRevResearch.2.043167}%
  \BibitemOpen
  \bibfield  {author} {\bibinfo {author} {\bibfnamefont {C.-H.}\ \bibnamefont
  {Liu}}, \bibinfo {author} {\bibfnamefont {K.}~\bibnamefont {Zhang}}, \bibinfo
  {author} {\bibfnamefont {Z.}~\bibnamefont {Yang}}, \ and\ \bibinfo {author}
  {\bibfnamefont {S.}~\bibnamefont {Chen}},\ }\href {\doibase
  10.1103/PhysRevResearch.2.043167} {\bibfield  {journal} {\bibinfo  {journal}
  {Phys. Rev. Research}\ }\textbf {\bibinfo {volume} {2}},\ \bibinfo {pages}
  {043167} (\bibinfo {year} {2020}{\natexlab{a}})}\BibitemShut {NoStop}%
\bibitem [{\citenamefont {Gong}\ \emph {et~al.}(2018)\citenamefont {Gong},
  \citenamefont {Ashida}, \citenamefont {Kawabata}, \citenamefont {Takasan},
  \citenamefont {Higashikawa},\ and\ \citenamefont {Ueda}}]{PhysRevX.8.031079}%
  \BibitemOpen
  \bibfield  {author} {\bibinfo {author} {\bibfnamefont {Z.}~\bibnamefont
  {Gong}}, \bibinfo {author} {\bibfnamefont {Y.}~\bibnamefont {Ashida}},
  \bibinfo {author} {\bibfnamefont {K.}~\bibnamefont {Kawabata}}, \bibinfo
  {author} {\bibfnamefont {K.}~\bibnamefont {Takasan}}, \bibinfo {author}
  {\bibfnamefont {S.}~\bibnamefont {Higashikawa}}, \ and\ \bibinfo {author}
  {\bibfnamefont {M.}~\bibnamefont {Ueda}},\ }\href {\doibase
  10.1103/PhysRevX.8.031079} {\bibfield  {journal} {\bibinfo  {journal} {Phys.
  Rev. X}\ }\textbf {\bibinfo {volume} {8}},\ \bibinfo {pages} {031079}
  (\bibinfo {year} {2018})}\BibitemShut {NoStop}%
\bibitem [{\citenamefont {Kawabata}\ \emph {et~al.}(2019)\citenamefont
  {Kawabata}, \citenamefont {Shiozaki}, \citenamefont {Ueda},\ and\
  \citenamefont {Sato}}]{PhysRevX.9.041015}%
  \BibitemOpen
  \bibfield  {author} {\bibinfo {author} {\bibfnamefont {K.}~\bibnamefont
  {Kawabata}}, \bibinfo {author} {\bibfnamefont {K.}~\bibnamefont {Shiozaki}},
  \bibinfo {author} {\bibfnamefont {M.}~\bibnamefont {Ueda}}, \ and\ \bibinfo
  {author} {\bibfnamefont {M.}~\bibnamefont {Sato}},\ }\href {\doibase
  10.1103/PhysRevX.9.041015} {\bibfield  {journal} {\bibinfo  {journal} {Phys.
  Rev. X}\ }\textbf {\bibinfo {volume} {9}},\ \bibinfo {pages} {041015}
  (\bibinfo {year} {2019})}\BibitemShut {NoStop}%
\bibitem [{\citenamefont {Okuma}\ and\ \citenamefont
  {Sato}(2019)}]{PhysRevLett.123.097701}%
  \BibitemOpen
  \bibfield  {author} {\bibinfo {author} {\bibfnamefont {N.}~\bibnamefont
  {Okuma}}\ and\ \bibinfo {author} {\bibfnamefont {M.}~\bibnamefont {Sato}},\
  }\href {\doibase 10.1103/PhysRevLett.123.097701} {\bibfield  {journal}
  {\bibinfo  {journal} {Phys. Rev. Lett.}\ }\textbf {\bibinfo {volume} {123}},\
  \bibinfo {pages} {097701} (\bibinfo {year} {2019})}\BibitemShut {NoStop}%
\bibitem [{\citenamefont {Okuma}\ \emph {et~al.}()\citenamefont {Okuma},
  \citenamefont {Kawabata}, \citenamefont {Shiozaki},\ and\ \citenamefont
  {Sato}}]{okumaTopologicalOriginNonHermitian2019}%
  \BibitemOpen
  \bibfield  {author} {\bibinfo {author} {\bibfnamefont {N.}~\bibnamefont
  {Okuma}}, \bibinfo {author} {\bibfnamefont {K.}~\bibnamefont {Kawabata}},
  \bibinfo {author} {\bibfnamefont {K.}~\bibnamefont {Shiozaki}}, \ and\
  \bibinfo {author} {\bibfnamefont {M.}~\bibnamefont {Sato}},\ }\href@noop {}
  {\ }\Eprint {http://arxiv.org/abs/1910.02878} {arXiv:1910.02878} \BibitemShut
  {NoStop}%
\bibitem [{\citenamefont {Kawabata}\ \emph
  {et~al.}(2020{\natexlab{a}})\citenamefont {Kawabata}, \citenamefont {Okuma},\
  and\ \citenamefont {Sato}}]{PhysRevB.101.195147}%
  \BibitemOpen
  \bibfield  {author} {\bibinfo {author} {\bibfnamefont {K.}~\bibnamefont
  {Kawabata}}, \bibinfo {author} {\bibfnamefont {N.}~\bibnamefont {Okuma}}, \
  and\ \bibinfo {author} {\bibfnamefont {M.}~\bibnamefont {Sato}},\ }\href
  {\doibase 10.1103/PhysRevB.101.195147} {\bibfield  {journal} {\bibinfo
  {journal} {Phys. Rev. B}\ }\textbf {\bibinfo {volume} {101}},\ \bibinfo
  {pages} {195147} (\bibinfo {year} {2020}{\natexlab{a}})}\BibitemShut
  {NoStop}%
\bibitem [{\citenamefont {Bessho}\ and\ \citenamefont
  {Sato}(2020)}]{bessho2020topological}%
  \BibitemOpen
  \bibfield  {author} {\bibinfo {author} {\bibfnamefont {T.}~\bibnamefont
  {Bessho}}\ and\ \bibinfo {author} {\bibfnamefont {M.}~\bibnamefont {Sato}},\
  }\href@noop {} {\  (\bibinfo {year} {2020})},\ \Eprint
  {http://arxiv.org/abs/2006.04204} {arXiv:2006.04204} \BibitemShut {NoStop}%
\bibitem [{\citenamefont {Okuma}\ and\ \citenamefont
  {Sato}(2020{\natexlab{a}})}]{okuma2020nonhermitian}%
  \BibitemOpen
  \bibfield  {author} {\bibinfo {author} {\bibfnamefont {N.}~\bibnamefont
  {Okuma}}\ and\ \bibinfo {author} {\bibfnamefont {M.}~\bibnamefont {Sato}},\
  }\href@noop {} {\  (\bibinfo {year} {2020}{\natexlab{a}})},\ \Eprint
  {http://arxiv.org/abs/2008.06498} {arXiv:2008.06498} \BibitemShut {NoStop}%
\bibitem [{\citenamefont {Kawabata}\ \emph
  {et~al.}(2020{\natexlab{b}})\citenamefont {Kawabata}, \citenamefont {Sato},\
  and\ \citenamefont {Shiozaki}}]{PhysRevB.102.205118}%
  \BibitemOpen
  \bibfield  {author} {\bibinfo {author} {\bibfnamefont {K.}~\bibnamefont
  {Kawabata}}, \bibinfo {author} {\bibfnamefont {M.}~\bibnamefont {Sato}}, \
  and\ \bibinfo {author} {\bibfnamefont {K.}~\bibnamefont {Shiozaki}},\ }\href
  {\doibase 10.1103/PhysRevB.102.205118} {\bibfield  {journal} {\bibinfo
  {journal} {Phys. Rev. B}\ }\textbf {\bibinfo {volume} {102}},\ \bibinfo
  {pages} {205118} (\bibinfo {year} {2020}{\natexlab{b}})}\BibitemShut
  {NoStop}%
\bibitem [{\citenamefont {Okuma}\ and\ \citenamefont
  {Sato}(2020{\natexlab{b}})}]{okuma2020quantum}%
  \BibitemOpen
  \bibfield  {author} {\bibinfo {author} {\bibfnamefont {N.}~\bibnamefont
  {Okuma}}\ and\ \bibinfo {author} {\bibfnamefont {M.}~\bibnamefont {Sato}},\
  }\href@noop {} {\  (\bibinfo {year} {2020}{\natexlab{b}})},\ \Eprint
  {http://arxiv.org/abs/2011.08175} {arXiv:2011.08175} \BibitemShut {NoStop}%
\bibitem [{\citenamefont {Kawabata}\ \emph
  {et~al.}(2020{\natexlab{c}})\citenamefont {Kawabata}, \citenamefont
  {Shiozaki},\ and\ \citenamefont {Ryu}}]{kawabata2020topological}%
  \BibitemOpen
  \bibfield  {author} {\bibinfo {author} {\bibfnamefont {K.}~\bibnamefont
  {Kawabata}}, \bibinfo {author} {\bibfnamefont {K.}~\bibnamefont {Shiozaki}},
  \ and\ \bibinfo {author} {\bibfnamefont {S.}~\bibnamefont {Ryu}},\
  }\href@noop {} {\  (\bibinfo {year} {2020}{\natexlab{c}})},\ \Eprint
  {http://arxiv.org/abs/2011.11449} {arXiv:2011.11449} \BibitemShut {NoStop}%
\bibitem [{\citenamefont {Lee}\ and\ \citenamefont
  {Thomale}(2019)}]{leeAnatomySkinModes2019b}%
  \BibitemOpen
  \bibfield  {author} {\bibinfo {author} {\bibfnamefont {C.~H.}\ \bibnamefont
  {Lee}}\ and\ \bibinfo {author} {\bibfnamefont {R.}~\bibnamefont {Thomale}},\
  }\href {\doibase 10.1103/PhysRevB.99.201103} {\bibfield  {journal} {\bibinfo
  {journal} {Phys. Rev. B}\ }\textbf {\bibinfo {volume} {99}},\ \bibinfo
  {pages} {201103} (\bibinfo {year} {2019})}\BibitemShut {NoStop}%
\bibitem [{\citenamefont {Lee}\ \emph {et~al.}(2019{\natexlab{a}})\citenamefont
  {Lee}, \citenamefont {Li},\ and\ \citenamefont
  {Gong}}]{PhysRevLett.123.016805}%
  \BibitemOpen
  \bibfield  {author} {\bibinfo {author} {\bibfnamefont {C.~H.}\ \bibnamefont
  {Lee}}, \bibinfo {author} {\bibfnamefont {L.}~\bibnamefont {Li}}, \ and\
  \bibinfo {author} {\bibfnamefont {J.}~\bibnamefont {Gong}},\ }\href {\doibase
  10.1103/PhysRevLett.123.016805} {\bibfield  {journal} {\bibinfo  {journal}
  {Phys. Rev. Lett.}\ }\textbf {\bibinfo {volume} {123}},\ \bibinfo {pages}
  {016805} (\bibinfo {year} {2019}{\natexlab{a}})}\BibitemShut {NoStop}%
\bibitem [{\citenamefont {Hofmann}\ \emph {et~al.}(2020)\citenamefont
  {Hofmann}, \citenamefont {Helbig}, \citenamefont {Schindler}, \citenamefont
  {Salgo}, \citenamefont {Brzezi\ifmmode~\acute{n}\else \'{n}\fi{}ska},
  \citenamefont {Greiter}, \citenamefont {Kiessling}, \citenamefont {Wolf},
  \citenamefont {Vollhardt}, \citenamefont {Kaba\ifmmode~\check{s}\else
  \v{s}\fi{}i}, \citenamefont {Lee}, \citenamefont {Bilu\ifmmode \check{s}\else
  \v{s}\fi{}i\ifmmode~\acute{c}\else \'{c}\fi{}}, \citenamefont {Thomale},\
  and\ \citenamefont {Neupert}}]{PhysRevResearch.2.023265}%
  \BibitemOpen
  \bibfield  {author} {\bibinfo {author} {\bibfnamefont {T.}~\bibnamefont
  {Hofmann}}, \bibinfo {author} {\bibfnamefont {T.}~\bibnamefont {Helbig}},
  \bibinfo {author} {\bibfnamefont {F.}~\bibnamefont {Schindler}}, \bibinfo
  {author} {\bibfnamefont {N.}~\bibnamefont {Salgo}}, \bibinfo {author}
  {\bibfnamefont {M.}~\bibnamefont {Brzezi\ifmmode~\acute{n}\else
  \'{n}\fi{}ska}}, \bibinfo {author} {\bibfnamefont {M.}~\bibnamefont
  {Greiter}}, \bibinfo {author} {\bibfnamefont {T.}~\bibnamefont {Kiessling}},
  \bibinfo {author} {\bibfnamefont {D.}~\bibnamefont {Wolf}}, \bibinfo {author}
  {\bibfnamefont {A.}~\bibnamefont {Vollhardt}}, \bibinfo {author}
  {\bibfnamefont {A.}~\bibnamefont {Kaba\ifmmode~\check{s}\else \v{s}\fi{}i}},
  \bibinfo {author} {\bibfnamefont {C.~H.}\ \bibnamefont {Lee}}, \bibinfo
  {author} {\bibfnamefont {A.}~\bibnamefont {Bilu\ifmmode \check{s}\else
  \v{s}\fi{}i\ifmmode~\acute{c}\else \'{c}\fi{}}}, \bibinfo {author}
  {\bibfnamefont {R.}~\bibnamefont {Thomale}}, \ and\ \bibinfo {author}
  {\bibfnamefont {T.}~\bibnamefont {Neupert}},\ }\href {\doibase
  10.1103/PhysRevResearch.2.023265} {\bibfield  {journal} {\bibinfo  {journal}
  {Phys. Rev. Research}\ }\textbf {\bibinfo {volume} {2}},\ \bibinfo {pages}
  {023265} (\bibinfo {year} {2020})}\BibitemShut {NoStop}%
\bibitem [{\citenamefont {Li}\ \emph {et~al.}(2020{\natexlab{a}})\citenamefont
  {Li}, \citenamefont {Lee},\ and\ \citenamefont
  {Gong}}]{PhysRevLett.124.250402}%
  \BibitemOpen
  \bibfield  {author} {\bibinfo {author} {\bibfnamefont {L.}~\bibnamefont
  {Li}}, \bibinfo {author} {\bibfnamefont {C.~H.}\ \bibnamefont {Lee}}, \ and\
  \bibinfo {author} {\bibfnamefont {J.}~\bibnamefont {Gong}},\ }\href {\doibase
  10.1103/PhysRevLett.124.250402} {\bibfield  {journal} {\bibinfo  {journal}
  {Phys. Rev. Lett.}\ }\textbf {\bibinfo {volume} {124}},\ \bibinfo {pages}
  {250402} (\bibinfo {year} {2020}{\natexlab{a}})}\BibitemShut {NoStop}%
\bibitem [{\citenamefont {Lee}\ \emph {et~al.}(2020)\citenamefont {Lee},
  \citenamefont {Li}, \citenamefont {Thomale},\ and\ \citenamefont
  {Gong}}]{PhysRevB.102.085151}%
  \BibitemOpen
  \bibfield  {author} {\bibinfo {author} {\bibfnamefont {C.~H.}\ \bibnamefont
  {Lee}}, \bibinfo {author} {\bibfnamefont {L.}~\bibnamefont {Li}}, \bibinfo
  {author} {\bibfnamefont {R.}~\bibnamefont {Thomale}}, \ and\ \bibinfo
  {author} {\bibfnamefont {J.}~\bibnamefont {Gong}},\ }\href {\doibase
  10.1103/PhysRevB.102.085151} {\bibfield  {journal} {\bibinfo  {journal}
  {Phys. Rev. B}\ }\textbf {\bibinfo {volume} {102}},\ \bibinfo {pages}
  {085151} (\bibinfo {year} {2020})}\BibitemShut {NoStop}%
\bibitem [{\citenamefont {Li}\ \emph {et~al.}(2020{\natexlab{b}})\citenamefont
  {Li}, \citenamefont {Lee}, \citenamefont {Mu},\ and\ \citenamefont
  {Gong}}]{Li:2020aa}%
  \BibitemOpen
  \bibfield  {author} {\bibinfo {author} {\bibfnamefont {L.}~\bibnamefont
  {Li}}, \bibinfo {author} {\bibfnamefont {C.~H.}\ \bibnamefont {Lee}},
  \bibinfo {author} {\bibfnamefont {S.}~\bibnamefont {Mu}}, \ and\ \bibinfo
  {author} {\bibfnamefont {J.}~\bibnamefont {Gong}},\ }\href {\doibase
  10.1038/s41467-020-18917-4} {\bibfield  {journal} {\bibinfo  {journal}
  {Nature Communications}\ }\textbf {\bibinfo {volume} {11}},\ \bibinfo {pages}
  {5491} (\bibinfo {year} {2020}{\natexlab{b}})}\BibitemShut {NoStop}%
\bibitem [{\citenamefont {Lee}\ and\ \citenamefont
  {Longhi}(2020)}]{Lee:2020aa}%
  \BibitemOpen
  \bibfield  {author} {\bibinfo {author} {\bibfnamefont {C.~H.}\ \bibnamefont
  {Lee}}\ and\ \bibinfo {author} {\bibfnamefont {S.}~\bibnamefont {Longhi}},\
  }\href {\doibase 10.1038/s42005-020-00417-y} {\bibfield  {journal} {\bibinfo
  {journal} {Communications Physics}\ }\textbf {\bibinfo {volume} {3}},\
  \bibinfo {pages} {147} (\bibinfo {year} {2020})}\BibitemShut {NoStop}%
\bibitem [{\citenamefont {Lee}(2020{\natexlab{a}})}]{lee2020manybody}%
  \BibitemOpen
  \bibfield  {author} {\bibinfo {author} {\bibfnamefont {C.~H.}\ \bibnamefont
  {Lee}},\ }\href@noop {} {\  (\bibinfo {year} {2020}{\natexlab{a}})},\ \Eprint
  {http://arxiv.org/abs/2006.01182} {arXiv:2006.01182} \BibitemShut {NoStop}%
\bibitem [{\citenamefont {Li}\ \emph {et~al.}(2020{\natexlab{c}})\citenamefont
  {Li}, \citenamefont {Lee},\ and\ \citenamefont {Gong}}]{li2020impurity}%
  \BibitemOpen
  \bibfield  {author} {\bibinfo {author} {\bibfnamefont {L.}~\bibnamefont
  {Li}}, \bibinfo {author} {\bibfnamefont {C.~H.}\ \bibnamefont {Lee}}, \ and\
  \bibinfo {author} {\bibfnamefont {J.}~\bibnamefont {Gong}},\ }\href@noop {}
  {\  (\bibinfo {year} {2020}{\natexlab{c}})},\ \Eprint
  {http://arxiv.org/abs/2008.05501} {arXiv:2008.05501} \BibitemShut {NoStop}%
\bibitem [{\citenamefont {Arouca}\ \emph {et~al.}(2020)\citenamefont {Arouca},
  \citenamefont {Lee},\ and\ \citenamefont {Smith}}]{arouca2020unconventional}%
  \BibitemOpen
  \bibfield  {author} {\bibinfo {author} {\bibfnamefont {R.}~\bibnamefont
  {Arouca}}, \bibinfo {author} {\bibfnamefont {C.~H.}\ \bibnamefont {Lee}}, \
  and\ \bibinfo {author} {\bibfnamefont {C.~M.}\ \bibnamefont {Smith}},\
  }\href@noop {} {\  (\bibinfo {year} {2020})},\ \Eprint
  {http://arxiv.org/abs/2009.03541} {arXiv:2009.03541} \BibitemShut {NoStop}%
\bibitem [{\citenamefont {Lee}(2020{\natexlab{b}})}]{lee2020exceptional}%
  \BibitemOpen
  \bibfield  {author} {\bibinfo {author} {\bibfnamefont {C.~H.}\ \bibnamefont
  {Lee}},\ }\href@noop {} {\  (\bibinfo {year} {2020}{\natexlab{b}})},\ \Eprint
  {http://arxiv.org/abs/2011.09505} {arXiv:2011.09505} \BibitemShut {NoStop}%
\bibitem [{\citenamefont {Pan}\ \emph {et~al.}(2020)\citenamefont {Pan},
  \citenamefont {Li},\ and\ \citenamefont {Gong}}]{pan2020pointgap}%
  \BibitemOpen
  \bibfield  {author} {\bibinfo {author} {\bibfnamefont {J.-S.}\ \bibnamefont
  {Pan}}, \bibinfo {author} {\bibfnamefont {L.}~\bibnamefont {Li}}, \ and\
  \bibinfo {author} {\bibfnamefont {J.}~\bibnamefont {Gong}},\ }\href@noop {}
  {\  (\bibinfo {year} {2020})},\ \Eprint {http://arxiv.org/abs/2010.14862}
  {arXiv:2010.14862} \BibitemShut {NoStop}%
\bibitem [{\citenamefont
  {Longhi}(2019{\natexlab{a}})}]{PhysRevLett.122.237601}%
  \BibitemOpen
  \bibfield  {author} {\bibinfo {author} {\bibfnamefont {S.}~\bibnamefont
  {Longhi}},\ }\href {\doibase 10.1103/PhysRevLett.122.237601} {\bibfield
  {journal} {\bibinfo  {journal} {Phys. Rev. Lett.}\ }\textbf {\bibinfo
  {volume} {122}},\ \bibinfo {pages} {237601} (\bibinfo {year}
  {2019}{\natexlab{a}})}\BibitemShut {NoStop}%
\bibitem [{\citenamefont
  {Longhi}(2019{\natexlab{b}})}]{PhysRevResearch.1.023013}%
  \BibitemOpen
  \bibfield  {author} {\bibinfo {author} {\bibfnamefont {S.}~\bibnamefont
  {Longhi}},\ }\href {\doibase 10.1103/PhysRevResearch.1.023013} {\bibfield
  {journal} {\bibinfo  {journal} {Phys. Rev. Research}\ }\textbf {\bibinfo
  {volume} {1}},\ \bibinfo {pages} {023013} (\bibinfo {year}
  {2019}{\natexlab{b}})}\BibitemShut {NoStop}%
\bibitem [{\citenamefont
  {Longhi}(2020{\natexlab{a}})}]{PhysRevLett.124.066602}%
  \BibitemOpen
  \bibfield  {author} {\bibinfo {author} {\bibfnamefont {S.}~\bibnamefont
  {Longhi}},\ }\href {\doibase 10.1103/PhysRevLett.124.066602} {\bibfield
  {journal} {\bibinfo  {journal} {Phys. Rev. Lett.}\ }\textbf {\bibinfo
  {volume} {124}},\ \bibinfo {pages} {066602} (\bibinfo {year}
  {2020}{\natexlab{a}})}\BibitemShut {NoStop}%
\bibitem [{\citenamefont {Longhi}(2020{\natexlab{b}})}]{longhi2020stochastic}%
  \BibitemOpen
  \bibfield  {author} {\bibinfo {author} {\bibfnamefont {S.}~\bibnamefont
  {Longhi}},\ }\href@noop {} {\  (\bibinfo {year} {2020}{\natexlab{b}})},\
  \Eprint {http://arxiv.org/abs/2008.06470} {arXiv:2008.06470} \BibitemShut
  {NoStop}%
\bibitem [{\citenamefont {Longhi}(2020{\natexlab{c}})}]{PhysRevB.102.201103}%
  \BibitemOpen
  \bibfield  {author} {\bibinfo {author} {\bibfnamefont {S.}~\bibnamefont
  {Longhi}},\ }\href {\doibase 10.1103/PhysRevB.102.201103} {\bibfield
  {journal} {\bibinfo  {journal} {Phys. Rev. B}\ }\textbf {\bibinfo {volume}
  {102}},\ \bibinfo {pages} {201103} (\bibinfo {year}
  {2020}{\natexlab{c}})}\BibitemShut {NoStop}%
\bibitem [{\citenamefont {Jiang}\ \emph {et~al.}(2019)\citenamefont {Jiang},
  \citenamefont {Lang}, \citenamefont {Yang}, \citenamefont {Zhu},\ and\
  \citenamefont {Chen}}]{PhysRevB.100.054301}%
  \BibitemOpen
  \bibfield  {author} {\bibinfo {author} {\bibfnamefont {H.}~\bibnamefont
  {Jiang}}, \bibinfo {author} {\bibfnamefont {L.-J.}\ \bibnamefont {Lang}},
  \bibinfo {author} {\bibfnamefont {C.}~\bibnamefont {Yang}}, \bibinfo {author}
  {\bibfnamefont {S.-L.}\ \bibnamefont {Zhu}}, \ and\ \bibinfo {author}
  {\bibfnamefont {S.}~\bibnamefont {Chen}},\ }\href {\doibase
  10.1103/PhysRevB.100.054301} {\bibfield  {journal} {\bibinfo  {journal}
  {Phys. Rev. B}\ }\textbf {\bibinfo {volume} {100}},\ \bibinfo {pages}
  {054301} (\bibinfo {year} {2019})}\BibitemShut {NoStop}%
\bibitem [{\citenamefont {Liu}\ \emph {et~al.}(2020{\natexlab{b}})\citenamefont
  {Liu}, \citenamefont {Zhou},\ and\ \citenamefont
  {Chen}}]{liu2020localization}%
  \BibitemOpen
  \bibfield  {author} {\bibinfo {author} {\bibfnamefont {Y.}~\bibnamefont
  {Liu}}, \bibinfo {author} {\bibfnamefont {Q.}~\bibnamefont {Zhou}}, \ and\
  \bibinfo {author} {\bibfnamefont {S.}~\bibnamefont {Chen}},\ }\href@noop {}
  {\  (\bibinfo {year} {2020}{\natexlab{b}})},\ \Eprint
  {http://arxiv.org/abs/2009.07605} {arXiv:2009.07605} \BibitemShut {NoStop}%
\bibitem [{\citenamefont {Liu}\ and\ \citenamefont
  {Chen}(2020)}]{PhysRevB.102.075404}%
  \BibitemOpen
  \bibfield  {author} {\bibinfo {author} {\bibfnamefont {Y.}~\bibnamefont
  {Liu}}\ and\ \bibinfo {author} {\bibfnamefont {S.}~\bibnamefont {Chen}},\
  }\href {\doibase 10.1103/PhysRevB.102.075404} {\bibfield  {journal} {\bibinfo
   {journal} {Phys. Rev. B}\ }\textbf {\bibinfo {volume} {102}},\ \bibinfo
  {pages} {075404} (\bibinfo {year} {2020})}\BibitemShut {NoStop}%
\bibitem [{\citenamefont {Xiong}(2018)}]{xiongWhyDoesBulk2018c}%
  \BibitemOpen
  \bibfield  {author} {\bibinfo {author} {\bibfnamefont {Y.}~\bibnamefont
  {Xiong}},\ }\href {\doibase 10.1088/2399-6528/aab64a} {\bibfield  {journal}
  {\bibinfo  {journal} {J. Phys. Commun.}\ }\textbf {\bibinfo {volume} {2}},\
  \bibinfo {pages} {035043} (\bibinfo {year} {2018})}\BibitemShut {NoStop}%
\bibitem [{\citenamefont {Martinez~Alvarez}\ \emph {et~al.}(2018)\citenamefont
  {Martinez~Alvarez}, \citenamefont {Barrios~Vargas},\ and\ \citenamefont
  {Foa~Torres}}]{PhysRevB.97.121401}%
  \BibitemOpen
  \bibfield  {author} {\bibinfo {author} {\bibfnamefont {V.~M.}\ \bibnamefont
  {Martinez~Alvarez}}, \bibinfo {author} {\bibfnamefont {J.~E.}\ \bibnamefont
  {Barrios~Vargas}}, \ and\ \bibinfo {author} {\bibfnamefont {L.~E.~F.}\
  \bibnamefont {Foa~Torres}},\ }\href {\doibase 10.1103/PhysRevB.97.121401}
  {\bibfield  {journal} {\bibinfo  {journal} {Phys. Rev. B}\ }\textbf {\bibinfo
  {volume} {97}},\ \bibinfo {pages} {121401} (\bibinfo {year}
  {2018})}\BibitemShut {NoStop}%
\bibitem [{\citenamefont {Deng}\ and\ \citenamefont
  {Yi}(2019{\natexlab{a}})}]{dengNonBlochTopologicalInvariants2019a}%
  \BibitemOpen
  \bibfield  {author} {\bibinfo {author} {\bibfnamefont {T.-S.}\ \bibnamefont
  {Deng}}\ and\ \bibinfo {author} {\bibfnamefont {W.}~\bibnamefont {Yi}},\
  }\href {\doibase 10.1103/PhysRevB.100.035102} {\bibfield  {journal} {\bibinfo
   {journal} {Phys. Rev. B}\ }\textbf {\bibinfo {volume} {100}},\ \bibinfo
  {pages} {035102} (\bibinfo {year} {2019}{\natexlab{a}})}\BibitemShut
  {NoStop}%
\bibitem [{\citenamefont {Deng}\ and\ \citenamefont
  {Yi}(2019{\natexlab{b}})}]{PhysRevB.100.035102}%
  \BibitemOpen
  \bibfield  {author} {\bibinfo {author} {\bibfnamefont {T.-S.}\ \bibnamefont
  {Deng}}\ and\ \bibinfo {author} {\bibfnamefont {W.}~\bibnamefont {Yi}},\
  }\href {\doibase 10.1103/PhysRevB.100.035102} {\bibfield  {journal} {\bibinfo
   {journal} {Phys. Rev. B}\ }\textbf {\bibinfo {volume} {100}},\ \bibinfo
  {pages} {035102} (\bibinfo {year} {2019}{\natexlab{b}})}\BibitemShut
  {NoStop}%
\bibitem [{\citenamefont {Li}\ \emph {et~al.}(2020{\natexlab{d}})\citenamefont
  {Li}, \citenamefont {Zhang},\ and\ \citenamefont
  {Yi}}]{li2020twodimensional}%
  \BibitemOpen
  \bibfield  {author} {\bibinfo {author} {\bibfnamefont {T.}~\bibnamefont
  {Li}}, \bibinfo {author} {\bibfnamefont {Y.-S.}\ \bibnamefont {Zhang}}, \
  and\ \bibinfo {author} {\bibfnamefont {W.}~\bibnamefont {Yi}},\ }\href@noop
  {} {\  (\bibinfo {year} {2020}{\natexlab{d}})},\ \Eprint
  {http://arxiv.org/abs/2005.09474} {arXiv:2005.09474} \BibitemShut {NoStop}%
\bibitem [{\citenamefont {Wang}\ \emph {et~al.}(2020)\citenamefont {Wang},
  \citenamefont {Guo},\ and\ \citenamefont {Kou}}]{PhysRevB.101.121116}%
  \BibitemOpen
  \bibfield  {author} {\bibinfo {author} {\bibfnamefont {X.-R.}\ \bibnamefont
  {Wang}}, \bibinfo {author} {\bibfnamefont {C.-X.}\ \bibnamefont {Guo}}, \
  and\ \bibinfo {author} {\bibfnamefont {S.-P.}\ \bibnamefont {Kou}},\ }\href
  {\doibase 10.1103/PhysRevB.101.121116} {\bibfield  {journal} {\bibinfo
  {journal} {Phys. Rev. B}\ }\textbf {\bibinfo {volume} {101}},\ \bibinfo
  {pages} {121116} (\bibinfo {year} {2020})}\BibitemShut {NoStop}%
\bibitem [{\citenamefont {Zhang}\ \emph
  {et~al.}(2020{\natexlab{c}})\citenamefont {Zhang}, \citenamefont {Chen},
  \citenamefont {Zhang}, \citenamefont {Lang}, \citenamefont {Li},\ and\
  \citenamefont {Zhu}}]{PhysRevB.101.235150}%
  \BibitemOpen
  \bibfield  {author} {\bibinfo {author} {\bibfnamefont {D.-W.}\ \bibnamefont
  {Zhang}}, \bibinfo {author} {\bibfnamefont {Y.-L.}\ \bibnamefont {Chen}},
  \bibinfo {author} {\bibfnamefont {G.-Q.}\ \bibnamefont {Zhang}}, \bibinfo
  {author} {\bibfnamefont {L.-J.}\ \bibnamefont {Lang}}, \bibinfo {author}
  {\bibfnamefont {Z.}~\bibnamefont {Li}}, \ and\ \bibinfo {author}
  {\bibfnamefont {S.-L.}\ \bibnamefont {Zhu}},\ }\href {\doibase
  10.1103/PhysRevB.101.235150} {\bibfield  {journal} {\bibinfo  {journal}
  {Phys. Rev. B}\ }\textbf {\bibinfo {volume} {101}},\ \bibinfo {pages}
  {235150} (\bibinfo {year} {2020}{\natexlab{c}})}\BibitemShut {NoStop}%
\bibitem [{\citenamefont {Scheibner}\ \emph {et~al.}(2020)\citenamefont
  {Scheibner}, \citenamefont {Irvine},\ and\ \citenamefont
  {Vitelli}}]{PhysRevLett.125.118001}%
  \BibitemOpen
  \bibfield  {author} {\bibinfo {author} {\bibfnamefont {C.}~\bibnamefont
  {Scheibner}}, \bibinfo {author} {\bibfnamefont {W.~T.~M.}\ \bibnamefont
  {Irvine}}, \ and\ \bibinfo {author} {\bibfnamefont {V.}~\bibnamefont
  {Vitelli}},\ }\href {\doibase 10.1103/PhysRevLett.125.118001} {\bibfield
  {journal} {\bibinfo  {journal} {Phys. Rev. Lett.}\ }\textbf {\bibinfo
  {volume} {125}},\ \bibinfo {pages} {118001} (\bibinfo {year}
  {2020})}\BibitemShut {NoStop}%
\bibitem [{\citenamefont {Yoshida}\ \emph {et~al.}(2020)\citenamefont
  {Yoshida}, \citenamefont {Mizoguchi},\ and\ \citenamefont
  {Hatsugai}}]{PhysRevResearch.2.022062}%
  \BibitemOpen
  \bibfield  {author} {\bibinfo {author} {\bibfnamefont {T.}~\bibnamefont
  {Yoshida}}, \bibinfo {author} {\bibfnamefont {T.}~\bibnamefont {Mizoguchi}},
  \ and\ \bibinfo {author} {\bibfnamefont {Y.}~\bibnamefont {Hatsugai}},\
  }\href {\doibase 10.1103/PhysRevResearch.2.022062} {\bibfield  {journal}
  {\bibinfo  {journal} {Phys. Rev. Research}\ }\textbf {\bibinfo {volume}
  {2}},\ \bibinfo {pages} {022062} (\bibinfo {year} {2020})}\BibitemShut
  {NoStop}%
\bibitem [{\citenamefont {He}\ \emph {et~al.}(2020)\citenamefont {He},
  \citenamefont {Fu}, \citenamefont {Zhang},\ and\ \citenamefont
  {Zhu}}]{PhysRevA.102.023308}%
  \BibitemOpen
  \bibfield  {author} {\bibinfo {author} {\bibfnamefont {P.}~\bibnamefont
  {He}}, \bibinfo {author} {\bibfnamefont {J.-H.}\ \bibnamefont {Fu}}, \bibinfo
  {author} {\bibfnamefont {D.-W.}\ \bibnamefont {Zhang}}, \ and\ \bibinfo
  {author} {\bibfnamefont {S.-L.}\ \bibnamefont {Zhu}},\ }\href {\doibase
  10.1103/PhysRevA.102.023308} {\bibfield  {journal} {\bibinfo  {journal}
  {Phys. Rev. A}\ }\textbf {\bibinfo {volume} {102}},\ \bibinfo {pages}
  {023308} (\bibinfo {year} {2020})}\BibitemShut {NoStop}%
\bibitem [{\citenamefont {Cao}\ \emph {et~al.}(2020)\citenamefont {Cao},
  \citenamefont {Li},\ and\ \citenamefont {Yang}}]{cao2020nonhermitian}%
  \BibitemOpen
  \bibfield  {author} {\bibinfo {author} {\bibfnamefont {Y.}~\bibnamefont
  {Cao}}, \bibinfo {author} {\bibfnamefont {Y.}~\bibnamefont {Li}}, \ and\
  \bibinfo {author} {\bibfnamefont {X.}~\bibnamefont {Yang}},\ }\href@noop {}
  {\  (\bibinfo {year} {2020})},\ \Eprint {http://arxiv.org/abs/2007.13499}
  {arXiv:2007.13499} \BibitemShut {NoStop}%
\bibitem [{\citenamefont {Claes}\ and\ \citenamefont
  {Hughes}(2020)}]{claes2020skin}%
  \BibitemOpen
  \bibfield  {author} {\bibinfo {author} {\bibfnamefont {J.}~\bibnamefont
  {Claes}}\ and\ \bibinfo {author} {\bibfnamefont {T.~L.}\ \bibnamefont
  {Hughes}},\ }\href@noop {} {\  (\bibinfo {year} {2020})},\ \Eprint
  {http://arxiv.org/abs/2007.03738} {arXiv:2007.03738} \BibitemShut {NoStop}%
\bibitem [{\citenamefont {Ma}\ and\ \citenamefont
  {Hughes}(2020)}]{ma2020quantum}%
  \BibitemOpen
  \bibfield  {author} {\bibinfo {author} {\bibfnamefont {Y.}~\bibnamefont
  {Ma}}\ and\ \bibinfo {author} {\bibfnamefont {T.~L.}\ \bibnamefont
  {Hughes}},\ }\href@noop {} {\  (\bibinfo {year} {2020})},\ \Eprint
  {http://arxiv.org/abs/2008.02284} {arXiv:2008.02284} \BibitemShut {NoStop}%
\bibitem [{\citenamefont {Okugawa}\ \emph {et~al.}(2020)\citenamefont
  {Okugawa}, \citenamefont {Takahashi},\ and\ \citenamefont
  {Yokomizo}}]{okugawa2020secondorder}%
  \BibitemOpen
  \bibfield  {author} {\bibinfo {author} {\bibfnamefont {R.}~\bibnamefont
  {Okugawa}}, \bibinfo {author} {\bibfnamefont {R.}~\bibnamefont {Takahashi}},
  \ and\ \bibinfo {author} {\bibfnamefont {K.}~\bibnamefont {Yokomizo}},\
  }\href@noop {} {\  (\bibinfo {year} {2020})},\ \Eprint
  {http://arxiv.org/abs/2008.03721} {arXiv:2008.03721} \BibitemShut {NoStop}%
\bibitem [{\citenamefont {Mandal}\ \emph {et~al.}(2020)\citenamefont {Mandal},
  \citenamefont {Banerjee}, \citenamefont {Ostrovskaya},\ and\ \citenamefont
  {Liew}}]{PhysRevLett.125.123902}%
  \BibitemOpen
  \bibfield  {author} {\bibinfo {author} {\bibfnamefont {S.}~\bibnamefont
  {Mandal}}, \bibinfo {author} {\bibfnamefont {R.}~\bibnamefont {Banerjee}},
  \bibinfo {author} {\bibfnamefont {E.~A.}\ \bibnamefont {Ostrovskaya}}, \ and\
  \bibinfo {author} {\bibfnamefont {T.~C.~H.}\ \bibnamefont {Liew}},\ }\href
  {\doibase 10.1103/PhysRevLett.125.123902} {\bibfield  {journal} {\bibinfo
  {journal} {Phys. Rev. Lett.}\ }\textbf {\bibinfo {volume} {125}},\ \bibinfo
  {pages} {123902} (\bibinfo {year} {2020})}\BibitemShut {NoStop}%
\bibitem [{\citenamefont {Gao}\ \emph {et~al.}(2020)\citenamefont {Gao},
  \citenamefont {Willatzen},\ and\ \citenamefont
  {Christensen}}]{PhysRevLett.125.206402}%
  \BibitemOpen
  \bibfield  {author} {\bibinfo {author} {\bibfnamefont {P.}~\bibnamefont
  {Gao}}, \bibinfo {author} {\bibfnamefont {M.}~\bibnamefont {Willatzen}}, \
  and\ \bibinfo {author} {\bibfnamefont {J.}~\bibnamefont {Christensen}},\
  }\href {\doibase 10.1103/PhysRevLett.125.206402} {\bibfield  {journal}
  {\bibinfo  {journal} {Phys. Rev. Lett.}\ }\textbf {\bibinfo {volume} {125}},\
  \bibinfo {pages} {206402} (\bibinfo {year} {2020})}\BibitemShut {NoStop}%
\bibitem [{\citenamefont {Yu}\ and\ \citenamefont
  {Deng}(2020)}]{yu2020unsupervised}%
  \BibitemOpen
  \bibfield  {author} {\bibinfo {author} {\bibfnamefont {L.-W.}\ \bibnamefont
  {Yu}}\ and\ \bibinfo {author} {\bibfnamefont {D.-L.}\ \bibnamefont {Deng}},\
  }\href@noop {} {\  (\bibinfo {year} {2020})},\ \Eprint
  {http://arxiv.org/abs/2010.14516} {arXiv:2010.14516} \BibitemShut {NoStop}%
\bibitem [{\citenamefont {Yoshida}(2020)}]{yoshida2020rdmft}%
  \BibitemOpen
  \bibfield  {author} {\bibinfo {author} {\bibfnamefont {T.}~\bibnamefont
  {Yoshida}},\ }\href@noop {} {\  (\bibinfo {year} {2020})},\ \Eprint
  {http://arxiv.org/abs/2011.04379} {arXiv:2011.04379} \BibitemShut {NoStop}%
\bibitem [{\citenamefont {wen Xiao}(2009)}]{xiao2009theory}%
  \BibitemOpen
  \bibfield  {author} {\bibinfo {author} {\bibfnamefont {M.}~\bibnamefont {wen
  Xiao}},\ }\href@noop {} {\  (\bibinfo {year} {2009})},\ \Eprint
  {http://arxiv.org/abs/0908.0787} {arXiv:0908.0787 [math-ph]} \BibitemShut
  {NoStop}%
\bibitem [{\citenamefont {Kawaguchi}\ and\ \citenamefont
  {Ueda}(2012)}]{KAWAGUCHI2012253}%
  \BibitemOpen
  \bibfield  {author} {\bibinfo {author} {\bibfnamefont {Y.}~\bibnamefont
  {Kawaguchi}}\ and\ \bibinfo {author} {\bibfnamefont {M.}~\bibnamefont
  {Ueda}},\ }\href {\doibase https://doi.org/10.1016/j.physrep.2012.07.005}
  {\bibfield  {journal} {\bibinfo  {journal} {Physics Reports}\ }\textbf
  {\bibinfo {volume} {520}},\ \bibinfo {pages} {253 } (\bibinfo {year}
  {2012})},\ \bibinfo {note} {spinor Bose--Einstein condensates}\BibitemShut
  {NoStop}%
\bibitem [{\citenamefont {Alase}\ \emph {et~al.}(2017)\citenamefont {Alase},
  \citenamefont {Cobanera}, \citenamefont {Ortiz},\ and\ \citenamefont
  {Viola}}]{PhysRevB.96.195133}%
  \BibitemOpen
  \bibfield  {author} {\bibinfo {author} {\bibfnamefont {A.}~\bibnamefont
  {Alase}}, \bibinfo {author} {\bibfnamefont {E.}~\bibnamefont {Cobanera}},
  \bibinfo {author} {\bibfnamefont {G.}~\bibnamefont {Ortiz}}, \ and\ \bibinfo
  {author} {\bibfnamefont {L.}~\bibnamefont {Viola}},\ }\href {\doibase
  10.1103/PhysRevB.96.195133} {\bibfield  {journal} {\bibinfo  {journal} {Phys.
  Rev. B}\ }\textbf {\bibinfo {volume} {96}},\ \bibinfo {pages} {195133}
  (\bibinfo {year} {2017})}\BibitemShut {NoStop}%
\bibitem [{\citenamefont {Cobanera}\ \emph {et~al.}(2018)\citenamefont
  {Cobanera}, \citenamefont {Alase}, \citenamefont {Ortiz},\ and\ \citenamefont
  {Viola}}]{PhysRevB.98.245423}%
  \BibitemOpen
  \bibfield  {author} {\bibinfo {author} {\bibfnamefont {E.}~\bibnamefont
  {Cobanera}}, \bibinfo {author} {\bibfnamefont {A.}~\bibnamefont {Alase}},
  \bibinfo {author} {\bibfnamefont {G.}~\bibnamefont {Ortiz}}, \ and\ \bibinfo
  {author} {\bibfnamefont {L.}~\bibnamefont {Viola}},\ }\href {\doibase
  10.1103/PhysRevB.98.245423} {\bibfield  {journal} {\bibinfo  {journal} {Phys.
  Rev. B}\ }\textbf {\bibinfo {volume} {98}},\ \bibinfo {pages} {245423}
  (\bibinfo {year} {2018})}\BibitemShut {NoStop}%
\bibitem [{\citenamefont {Ozawa}\ \emph {et~al.}(2019)\citenamefont {Ozawa},
  \citenamefont {Price}, \citenamefont {Amo}, \citenamefont {Goldman},
  \citenamefont {Hafezi}, \citenamefont {Lu}, \citenamefont {Rechtsman},
  \citenamefont {Schuster}, \citenamefont {Simon}, \citenamefont {Zilberberg},\
  and\ \citenamefont {Carusotto}}]{RevModPhys.91.015006}%
  \BibitemOpen
  \bibfield  {author} {\bibinfo {author} {\bibfnamefont {T.}~\bibnamefont
  {Ozawa}}, \bibinfo {author} {\bibfnamefont {H.~M.}\ \bibnamefont {Price}},
  \bibinfo {author} {\bibfnamefont {A.}~\bibnamefont {Amo}}, \bibinfo {author}
  {\bibfnamefont {N.}~\bibnamefont {Goldman}}, \bibinfo {author} {\bibfnamefont
  {M.}~\bibnamefont {Hafezi}}, \bibinfo {author} {\bibfnamefont
  {L.}~\bibnamefont {Lu}}, \bibinfo {author} {\bibfnamefont {M.~C.}\
  \bibnamefont {Rechtsman}}, \bibinfo {author} {\bibfnamefont {D.}~\bibnamefont
  {Schuster}}, \bibinfo {author} {\bibfnamefont {J.}~\bibnamefont {Simon}},
  \bibinfo {author} {\bibfnamefont {O.}~\bibnamefont {Zilberberg}}, \ and\
  \bibinfo {author} {\bibfnamefont {I.}~\bibnamefont {Carusotto}},\ }\href
  {\doibase 10.1103/RevModPhys.91.015006} {\bibfield  {journal} {\bibinfo
  {journal} {Rev. Mod. Phys.}\ }\textbf {\bibinfo {volume} {91}},\ \bibinfo
  {pages} {015006} (\bibinfo {year} {2019})}\BibitemShut {NoStop}%
\bibitem [{\citenamefont {Matsumoto}\ \emph {et~al.}(2014)\citenamefont
  {Matsumoto}, \citenamefont {Shindou},\ and\ \citenamefont
  {Murakami}}]{PhysRevB.89.054420}%
  \BibitemOpen
  \bibfield  {author} {\bibinfo {author} {\bibfnamefont {R.}~\bibnamefont
  {Matsumoto}}, \bibinfo {author} {\bibfnamefont {R.}~\bibnamefont {Shindou}},
  \ and\ \bibinfo {author} {\bibfnamefont {S.}~\bibnamefont {Murakami}},\
  }\href {\doibase 10.1103/PhysRevB.89.054420} {\bibfield  {journal} {\bibinfo
  {journal} {Phys. Rev. B}\ }\textbf {\bibinfo {volume} {89}},\ \bibinfo
  {pages} {054420} (\bibinfo {year} {2014})}\BibitemShut {NoStop}%
\bibitem [{\citenamefont {Barnett}(2013)}]{PhysRevA.88.063631}%
  \BibitemOpen
  \bibfield  {author} {\bibinfo {author} {\bibfnamefont {R.}~\bibnamefont
  {Barnett}},\ }\href {\doibase 10.1103/PhysRevA.88.063631} {\bibfield
  {journal} {\bibinfo  {journal} {Phys. Rev. A}\ }\textbf {\bibinfo {volume}
  {88}},\ \bibinfo {pages} {063631} (\bibinfo {year} {2013})}\BibitemShut
  {NoStop}%
\bibitem [{\citenamefont {Ohashi}\ \emph {et~al.}(2020)\citenamefont {Ohashi},
  \citenamefont {Kobayashi},\ and\ \citenamefont
  {Kawaguchi}}]{PhysRevA.101.013625}%
  \BibitemOpen
  \bibfield  {author} {\bibinfo {author} {\bibfnamefont {T.}~\bibnamefont
  {Ohashi}}, \bibinfo {author} {\bibfnamefont {S.}~\bibnamefont {Kobayashi}}, \
  and\ \bibinfo {author} {\bibfnamefont {Y.}~\bibnamefont {Kawaguchi}},\ }\href
  {\doibase 10.1103/PhysRevA.101.013625} {\bibfield  {journal} {\bibinfo
  {journal} {Phys. Rev. A}\ }\textbf {\bibinfo {volume} {101}},\ \bibinfo
  {pages} {013625} (\bibinfo {year} {2020})}\BibitemShut {NoStop}%
\bibitem [{\citenamefont {Galilo}\ \emph {et~al.}(2015)\citenamefont {Galilo},
  \citenamefont {Lee},\ and\ \citenamefont {Barnett}}]{PhysRevLett.115.245302}%
  \BibitemOpen
  \bibfield  {author} {\bibinfo {author} {\bibfnamefont {B.}~\bibnamefont
  {Galilo}}, \bibinfo {author} {\bibfnamefont {D.~K.~K.}\ \bibnamefont {Lee}},
  \ and\ \bibinfo {author} {\bibfnamefont {R.}~\bibnamefont {Barnett}},\ }\href
  {\doibase 10.1103/PhysRevLett.115.245302} {\bibfield  {journal} {\bibinfo
  {journal} {Phys. Rev. Lett.}\ }\textbf {\bibinfo {volume} {115}},\ \bibinfo
  {pages} {245302} (\bibinfo {year} {2015})}\BibitemShut {NoStop}%
\bibitem [{\citenamefont {Engelhardt}\ \emph {et~al.}(2016)\citenamefont
  {Engelhardt}, \citenamefont {Benito}, \citenamefont {Platero},\ and\
  \citenamefont {Brandes}}]{PhysRevLett.117.045302}%
  \BibitemOpen
  \bibfield  {author} {\bibinfo {author} {\bibfnamefont {G.}~\bibnamefont
  {Engelhardt}}, \bibinfo {author} {\bibfnamefont {M.}~\bibnamefont {Benito}},
  \bibinfo {author} {\bibfnamefont {G.}~\bibnamefont {Platero}}, \ and\
  \bibinfo {author} {\bibfnamefont {T.}~\bibnamefont {Brandes}},\ }\href
  {\doibase 10.1103/PhysRevLett.117.045302} {\bibfield  {journal} {\bibinfo
  {journal} {Phys. Rev. Lett.}\ }\textbf {\bibinfo {volume} {117}},\ \bibinfo
  {pages} {045302} (\bibinfo {year} {2016})}\BibitemShut {NoStop}%
\bibitem [{\citenamefont {Peano}\ \emph {et~al.}(2016)\citenamefont {Peano},
  \citenamefont {Houde}, \citenamefont {Marquardt},\ and\ \citenamefont
  {Clerk}}]{PhysRevX.6.041026}%
  \BibitemOpen
  \bibfield  {author} {\bibinfo {author} {\bibfnamefont {V.}~\bibnamefont
  {Peano}}, \bibinfo {author} {\bibfnamefont {M.}~\bibnamefont {Houde}},
  \bibinfo {author} {\bibfnamefont {F.}~\bibnamefont {Marquardt}}, \ and\
  \bibinfo {author} {\bibfnamefont {A.~A.}\ \bibnamefont {Clerk}},\ }\href
  {\doibase 10.1103/PhysRevX.6.041026} {\bibfield  {journal} {\bibinfo
  {journal} {Phys. Rev. X}\ }\textbf {\bibinfo {volume} {6}},\ \bibinfo {pages}
  {041026} (\bibinfo {year} {2016})}\BibitemShut {NoStop}%
\bibitem [{\citenamefont {McDonald}\ \emph {et~al.}(2018)\citenamefont
  {McDonald}, \citenamefont {Pereg-Barnea},\ and\ \citenamefont
  {Clerk}}]{PhysRevX.8.041031}%
  \BibitemOpen
  \bibfield  {author} {\bibinfo {author} {\bibfnamefont {A.}~\bibnamefont
  {McDonald}}, \bibinfo {author} {\bibfnamefont {T.}~\bibnamefont
  {Pereg-Barnea}}, \ and\ \bibinfo {author} {\bibfnamefont {A.~A.}\
  \bibnamefont {Clerk}},\ }\href {\doibase 10.1103/PhysRevX.8.041031}
  {\bibfield  {journal} {\bibinfo  {journal} {Phys. Rev. X}\ }\textbf {\bibinfo
  {volume} {8}},\ \bibinfo {pages} {041031} (\bibinfo {year}
  {2018})}\BibitemShut {NoStop}%
\bibitem [{\citenamefont {Wu}\ and\ \citenamefont
  {Niu}(2001)}]{PhysRevA.64.061603}%
  \BibitemOpen
  \bibfield  {author} {\bibinfo {author} {\bibfnamefont {B.}~\bibnamefont
  {Wu}}\ and\ \bibinfo {author} {\bibfnamefont {Q.}~\bibnamefont {Niu}},\
  }\href {\doibase 10.1103/PhysRevA.64.061603} {\bibfield  {journal} {\bibinfo
  {journal} {Phys. Rev. A}\ }\textbf {\bibinfo {volume} {64}},\ \bibinfo
  {pages} {061603} (\bibinfo {year} {2001})}\BibitemShut {NoStop}%
\bibitem [{\citenamefont {Liu}\ \emph {et~al.}(2019)\citenamefont {Liu},
  \citenamefont {Jiang},\ and\ \citenamefont {Chen}}]{PhysRevB.99.125103}%
  \BibitemOpen
  \bibfield  {author} {\bibinfo {author} {\bibfnamefont {C.-H.}\ \bibnamefont
  {Liu}}, \bibinfo {author} {\bibfnamefont {H.}~\bibnamefont {Jiang}}, \ and\
  \bibinfo {author} {\bibfnamefont {S.}~\bibnamefont {Chen}},\ }\href {\doibase
  10.1103/PhysRevB.99.125103} {\bibfield  {journal} {\bibinfo  {journal} {Phys.
  Rev. B}\ }\textbf {\bibinfo {volume} {99}},\ \bibinfo {pages} {125103}
  (\bibinfo {year} {2019})}\BibitemShut {NoStop}%
\bibitem [{\citenamefont {Liu}\ and\ \citenamefont
  {Chen}(2019)}]{PhysRevB.100.144106}%
  \BibitemOpen
  \bibfield  {author} {\bibinfo {author} {\bibfnamefont {C.-H.}\ \bibnamefont
  {Liu}}\ and\ \bibinfo {author} {\bibfnamefont {S.}~\bibnamefont {Chen}},\
  }\href {\doibase 10.1103/PhysRevB.100.144106} {\bibfield  {journal} {\bibinfo
   {journal} {Phys. Rev. B}\ }\textbf {\bibinfo {volume} {100}},\ \bibinfo
  {pages} {144106} (\bibinfo {year} {2019})}\BibitemShut {NoStop}%
\bibitem [{\citenamefont {Zhou}\ and\ \citenamefont
  {Lee}(2019)}]{PhysRevB.99.235112}%
  \BibitemOpen
  \bibfield  {author} {\bibinfo {author} {\bibfnamefont {H.}~\bibnamefont
  {Zhou}}\ and\ \bibinfo {author} {\bibfnamefont {J.~Y.}\ \bibnamefont {Lee}},\
  }\href {\doibase 10.1103/PhysRevB.99.235112} {\bibfield  {journal} {\bibinfo
  {journal} {Phys. Rev. B}\ }\textbf {\bibinfo {volume} {99}},\ \bibinfo
  {pages} {235112} (\bibinfo {year} {2019})}\BibitemShut {NoStop}%
\bibitem [{\citenamefont {Lee}\ \emph {et~al.}(2019{\natexlab{b}})\citenamefont
  {Lee}, \citenamefont {Ahn}, \citenamefont {Zhou},\ and\ \citenamefont
  {Vishwanath}}]{PhysRevLett.123.206404}%
  \BibitemOpen
  \bibfield  {author} {\bibinfo {author} {\bibfnamefont {J.~Y.}\ \bibnamefont
  {Lee}}, \bibinfo {author} {\bibfnamefont {J.}~\bibnamefont {Ahn}}, \bibinfo
  {author} {\bibfnamefont {H.}~\bibnamefont {Zhou}}, \ and\ \bibinfo {author}
  {\bibfnamefont {A.}~\bibnamefont {Vishwanath}},\ }\href {\doibase
  10.1103/PhysRevLett.123.206404} {\bibfield  {journal} {\bibinfo  {journal}
  {Phys. Rev. Lett.}\ }\textbf {\bibinfo {volume} {123}},\ \bibinfo {pages}
  {206404} (\bibinfo {year} {2019}{\natexlab{b}})}\BibitemShut {NoStop}%
\bibitem [{\citenamefont {Brody}(2013)}]{Brody_2013}%
  \BibitemOpen
  \bibfield  {author} {\bibinfo {author} {\bibfnamefont {D.~C.}\ \bibnamefont
  {Brody}},\ }\href {\doibase 10.1088/1751-8113/47/3/035305} {\bibfield
  {journal} {\bibinfo  {journal} {Journal of Physics A: Mathematical and
  Theoretical}\ }\textbf {\bibinfo {volume} {47}},\ \bibinfo {pages} {035305}
  (\bibinfo {year} {2013})}\BibitemShut {NoStop}%
\bibitem [{\citenamefont {Yokomizo}\ and\ \citenamefont
  {Murakami}(2020{\natexlab{b}})}]{yokomizo2020nonbloch}%
  \BibitemOpen
  \bibfield  {author} {\bibinfo {author} {\bibfnamefont {K.}~\bibnamefont
  {Yokomizo}}\ and\ \bibinfo {author} {\bibfnamefont {S.}~\bibnamefont
  {Murakami}},\ }\href@noop {} {\  (\bibinfo {year} {2020}{\natexlab{b}})},\
  \Eprint {http://arxiv.org/abs/2012.00439} {arXiv:2012.00439
  [cond-mat.mes-hall]} \BibitemShut {NoStop}%
\end{thebibliography}%
\bibliographystyle{apsrev4-1}

\end{document}